\documentclass[3p,number,preprint]{elsarticle}

\usepackage[usenames,dvipsnames]{color}
\usepackage{algpseudocode}
\usepackage{algorithm}
\usepackage{amssymb}
\usepackage{amsmath}
\usepackage{multirow}
\usepackage{graphicx}
\usepackage{hyperref}
\usepackage{xspace}
\usepackage{subfig}
\usepackage{cases}

\journal{Physica A}

\DeclareGraphicsExtensions{.eps,.pdf,.jpg,.png}

\definecolor{black}{RGB}{0, 0, 0}
\definecolor{gray}{RGB}{121, 121, 121}
\definecolor{blue}{RGB}{0, 84, 147}
\definecolor{green}{RGB}{146, 144, 0}
\definecolor{mocha}{RGB}{147, 82, 0}
\definecolor{asparagus}{RGB}{146, 144, 0}

\newcommand{\argmax}[0]{\arg\!\max}
\newcommand{\mean}[1]{\left<#1\right>}

\newcommand{\links}[0]{L}
\newcommand{\nodes}[0]{V}
\newcommand{\graph}[0]{G}
\newcommand{\group}[0]{\mathcal{G}}
\newcommand{\hierarchy}[0]{\mathcal{H}}

\newcommand{\n}[0]{n}
\newcommand{\m}[0]{m}
\newcommand{\M}[0]{M}

\newcommand{\levels}[0]{L}
\newcommand{\random}[0]{p}
\newcommand{\mixing}[0]{\mu}
\newcommand{\probability}[0]{\theta}

\newcommand{\node}[0]{v}

\newcommand{\inner}[0]{\mathcal{I}}
\newcommand{\order}[0]{t}
\newcommand{\labeling}[0]{g}
\newcommand{\balance}[0]{b}
\newcommand{\diffusion}[0]{f}
\newcommand{\diffusiont}[0]{\diffusion'}

\newcommand{\degree}[0]{k}
\newcommand{\triads}[0]{\omega}
\newcommand{\triangles}[0]{\Delta}
\newcommand{\clustering}[0]{c}
\newcommand{\dclustering}[0]{d}
\newcommand{\CLUSTERING}[0]{C}
\newcommand{\DCLUSTERING}[0]{D}
\newcommand{\neighbors}[0]{\Gamma}

\newcommand{\complexity}[1]{\mathcal{O}(#1)}
\newcommand{\likelihood}[0]{\mathcal{L}}
\newcommand{\LOGL}[0]{$\log\likelihood$\xspace}
\newcommand{\MLOGL}[0]{$-\log\likelihood$\xspace}
\newcommand{\logl}[1]{$\log\likelihood=#1$\xspace}

\newcommand{\parameter}[0]{\nu}
\newcommand{\stability}[0]{\eta}

\newcommand{\LPA}[0]{\textit{LPA}\xspace}
\newcommand{\GMO}[0]{\textit{GMO}\xspace}
\newcommand{\LUV}[0]{\textit{LUV}\xspace}
\newcommand{\SCP}[0]{\textit{SCP}\xspace}
\newcommand{\MCL}[0]{\textit{MCL}\xspace}
\newcommand{\IMD}[0]{\textit{IMD}\xspace}
\newcommand{\IMP}[0]{\textit{IMP}\xspace}
\newcommand{\NMF}[0]{\textit{NMF}\xspace}
\newcommand{\KMN}[0]{\textit{KMN}\xspace}
\newcommand{\EMM}[0]{\textit{EMM}\xspace}
\newcommand{\DMM}[0]{\textit{DMM}\xspace}
\newcommand{\GPA}[0]{\textit{GPA}\xspace}
\newcommand{\MPA}[0]{\textit{MPA}\xspace}
\newcommand{\HPA}[0]{\textit{HPA}\xspace}

\newcommand{\LFR}[0]{\textit{LFR}\xspace}
\newcommand{\GN}[0]{\textit{GN}\xspace}
\newcommand{\GNT}[0]{\textit{GN2}\xspace}
\newcommand{\SBV}[0]{\textit{SB2}\xspace}
\newcommand{\SBX}[0]{\textit{SB3}\xspace}

\newcommand{\AFL}[0]{\textit{football}\xspace}
\newcommand{\NSC}[0]{\textit{science}\xspace}
\newcommand{\EUR}[0]{\textit{europe}\xspace}
\newcommand{\PLB}[0]{\textit{books}\xspace}
\newcommand{\CLT}[0]{\textit{compute}\xspace}
\newcommand{\JNG}[0]{\textit{jung}\xspace}
\newcommand{\JVX}[0]{\textit{javax}\xspace}
\newcommand{\MNE}[0]{\textit{elegans}\xspace}
\newcommand{\SWC}[0]{\textit{women}\xspace}
\newcommand{\SCI}[0]{\textit{corporate}\xspace}

\newcommand{\NMI}[0]{\textit{NMI}\xspace}
\newcommand{\NVI}[0]{\textit{NVI}\xspace}
\newcommand{\ARI}[0]{\textit{ARI}\xspace}
\newcommand{\AUC}[0]{\textit{AUC}\xspace}

\newcommand{\HLINE}[0]{\hline\noalign{\smallskip}}
\newcommand{\TLINE}[0]{\noalign{\smallskip}\HLINE}
\newcommand{\NULL}[0]{-\xspace}

\newcommand{\plotwidth}[0]{0.325\textwidth}
\newcommand{\smallwidth}[0]{0.30\textwidth}
\newcommand{\figurewidth}[0]{0.35\textwidth}

\algrenewcommand{\algorithmiccomment}[1]{\hfill\{#1\}}

\newcommand{\equref}[1]{Eq.~(\ref{equ:#1})\xspace}
\newcommand{\secref}[1]{Section~\ref{sec:#1}\xspace}
\renewcommand{\algref}[1]{Alg.~\ref{alg:#1}\xspace}
\newcommand{\figref}[1]{Fig.~\ref{fig:#1}\xspace}
\newcommand{\tblref}[1]{Table~\ref{tbl:#1}\xspace}

\begin{document}


\begin{frontmatter}

\title{Group detection in complex networks:\\An algorithm and comparison of the state-of-the-art}

\author{Lovro \v{S}ubelj\corref{coraut}}
\cortext[coraut]{Corresponding author. Tel.: +386 1 476 81 86.}
\ead{lovro.subelj@fri.uni-lj.si}

\author{Marko Bajec}
\ead{marko.bajec@fri.uni-lj.si}

\address{University of Ljubljana, Faculty of Computer and Information Science, Tr\v{z}a\v{s}ka 25, SI-1000 Ljubljana, Slovenia}


\begin{abstract}
Complex real-world networks commonly reveal characteristic groups of nodes like communities and modules. These are of value in various applications, especially in the case of large social and information networks. However, while numerous community detection techniques have been presented in the literature, approaches for other groups of nodes are relatively rare and often limited in some way. We present a simple propagation-based algorithm for general group detection that requires no apriori knowledge and has near ideal complexity. The main novelty here is that different types of groups are revealed through an adequate hierarchical group refinement procedure. The proposed algorithm is validated on various synthetic and real-world networks, and rigorously compared against twelve other state-of-the-art approaches on group detection, hierarchy discovery and link prediction tasks. The algorithm is comparable to the state-of-the-art in community detection, while superior in general group detection and link prediction. Based on the comparison, we also discuss some prominent directions for future work on group detection in complex networks.
\end{abstract}


\end{frontmatter}


\section{Introduction} \label{sec:intro}
Complex networks of real-world systems commonly reveal groups of nodes with characteristic connection pattern~\cite{New12} (e.g., densely connected groups known as communities~\cite{GN02}). These correspond to people with common interests in social networks~\cite{DDDA05} or classes with the same information signature in software networks~\cite{SB12u}. Characteristic groups of nodes provide an important insight into the structure and function of real-world networks~\cite{PDFV05}, while group detection also has numerous practical applications, including epidemic outbreak prevention~\cite{RZ07}, viral marketing~\cite{LAH07}, software package prediction~\cite{SB12s} and compression~\cite{BRSV11}.

Despite an outburst of community detection algorithms in the last decade~\cite{For10,New12}, approaches for other groups of nodes are relatively rare and often limited (e.g., demand some apriori knowledge about the network structure). Thus, we here propose a general group detection algorithm based on the label propagation framework in~\cite{SB12u,SB11g} that requires no apriori knowledge. Analysis in the paper confirms that the proposed algorithm is at least comparable to the current state-of-the-art, while its complexity is near ideal. The main novelty of the paper is else a simple hierarchical refinement procedure that enables straightforward discovery of different types of groups. The paper also includes a detailed empirical comparison of a larger number of state-of-the-art approaches for group detection that may be of an independent interest.

The rest of the paper is structured as follows. First, \secref{background} gives preliminary discussion on groups of nodes in real-world networks. Next, we introduce the group detection algorithm proposed in the paper in~\secref{algorithm}. Rigorous analysis on synthetic and real-world network appears in~\secref{analysis}, whereas detailed comparison with the state-of-the-art on group detection, hierarchy discovery and link prediction tasks is presented in~\secref{comparison}. \secref{conclusions} concludes the paper and gives some prominent directions for future work.


\section{Background} \label{sec:background}
Let the network be represented by a simple undirected graph $\graph(\nodes,\links)$, where $\nodes$ is a set of nodes in the network, $|\nodes|=\n$, and $\links$ is the set of links, $|\links|=\m$. Next, let $\neighbors_i$ be the set of neighbors of node $\node_i$, $\node_i\in\nodes$, and $\triangles_i$ the number of links between the nodes in $\neighbors_i$. Last, let $\degree_i$ be the degree of $\node_i$, $|\neighbors_i|=\degree_i$, and let $\mean{\degree}$ be the average degree in the network.

\subsection{Groups in real-world networks} \label{sec:background:groups}
The present paper is concerned with groups of nodes with characteristic connection pattern that appear in complex real-world networks~\cite{NL07}. While these could be defined in various ways, we adopt two types of groups that have been most popular in the recent literature~\cite{For10,New12}.

First, we consider communities~\cite{GN02} (also link-density community~\cite{LXZY07}) that, e.g., represent groups of people with common interests in social networks~\cite{DDDA05}. Community is defined as a (connected) group of nodes with more links towards the nodes in the group than to the rest of the network~\cite{RCCLP04}.

Second, we consider sparse modules~\cite{PSR10} (also link-pattern community~\cite{LXZY07} and other~\cite{RW07}) that in, e.g., software networks correspond to classes with the same function~\cite{SB12u}. Module is defined as a (possibly) disconnected group of nodes with more links towards common neighbors than to the rest of the network~\cite{SB12u}.

Definition of a module is rather similar to the concept of structural or regular equivalence~\cite{LW71,EB94} (i.e., blockmodels), although not equivalent. Note that communities can in fact be considered under the definition of modules (most authors have indeed adopted this stance~\cite{NL07,RW07}), however, there exist important differences between the two~\cite{SB12u} (e.g., connectedness). 

\subsection{Node and network clustering} \label{sec:background:clustering}
Based on the above, a group detection approach could gain from differentiating between communities and modules. Note that the definition of a community implies locally dense structure, while no such tendency appears in the definition of a module. Density of node's neighborhood is usually measured by the node clustering coefficient $\clustering$~\cite{WS98} defined~as
\begin{eqnarray}
\label{equ:clustering}
\clustering_i & = & \frac{\triangles_i}{\binom{\degree_i}{2}},
\end{eqnarray}
while $\CLUSTERING= \frac{1}{\n} \sum_i \clustering_i$ is the network clustering coefficient~\cite{WS98,NSW01}.

However, denominator in~\equref{clustering} introduces biases into the definition of clustering, since $\binom{\degree_i}{2}$ often cannot be reached due to fixed degree distribution~\cite{SV05}. The latter is particularly apparent in degree disassortative networks~\cite{New03a}, which most real-world networks in fact are. Thus, an alternative definition denoted degree-corrected clustering coefficient $\dclustering$~\cite{SV05} has been proposed as
\begin{eqnarray}
\label{equ:dclustering}
\dclustering_i & = & \frac{\triangles_i}{\triads_i},
\end{eqnarray}
where $\triads_i$ is the maximum possible number of links among~$\neighbors_i$ and $\DCLUSTERING=\frac{1}{\n} \sum_i \dclustering_i$.

Thus, in the presence of communities, one can expect dense network structure with $\DCLUSTERING\gg 0$. On the other hand, different configurations of modules (e.g., bipartite, multi-partite or star-like structures) imply sparser networks with $\DCLUSTERING\approx 0$. For example, most prominent modules are found in two-mode networks, where $\DCLUSTERING=0$. It should, however, be stressed that a module structure does not necessarily mean $\DCLUSTERING=0$, since a tripartite configuration of modules obviously has $\DCLUSTERING>0$.


\section{The algorithm} \label{sec:algorithm}
The proposed algorithm for group detection is based on the label propagation framework~\cite{SB12u,SB11g} that we introduce next. Due to simplicity, the framework is presented for the case of simple undirected graphs, although a generalization to weighted multigraphs is straightforward.

\subsection{Group detection by propagation} \label{sec:algorithm:gpa}
Let $\labeling_i$ be an unknown group label of node $\node_i$ and let $\group_i$ a group of nodes with label $\labeling_i$. Propagating labels between the nodes was first proposed for community detection~\cite{RAK07}. Initially, each node is labeled with a unique label as $\labeling_i=i$ (i.e., put in a separate group $\group_i$). Then, at each iteration of the algorithm, each node adopts the label shared by most of its neighbors (ties are broken uniformly at random). Hence,
\begin{eqnarray}
\label{equ:lpa}
\labeling_i & = & \argmax_\labeling \sum_{\node_j\in\neighbors_i} \delta(\labeling_j,\labeling),
\end{eqnarray}
where $\neighbors_i$ is the set of neighbors of node $\node_i$ and the $\delta$ is a Kronecker delta. Due to the presence of numerous links within the communities, relative to the number of links towards the rest of the network, nodes in communities form a consensus on some label $\labeling$ after only a few iterations. Thus, when the process converges, disconnected groups $\{\group\}$ are classified as communities. Due to very fast structural inference of label propagation, the algorithm exhibits near linear complexity, while the expected number of iterations on a network with a billion links is only $113$~\cite{SB11d}.

The approach can be significantly improved by adopting also node preferences~\cite{LHLC09}. Preferences adjust the propagation strength of the nodes and force the label propagation process towards a desirable group partition $\{\group\}$. Let~$\diffusion_i$ and $\balance_i$ be the preferences of node $\node_i$~\cite{SB12u} (see below). Then, \equref{lpa} is rewritten to
\begin{eqnarray}
\label{equ:dpa}
\labeling_i & = & \argmax_\labeling \sum_{\node_j\in\neighbors_i} \diffusion_j\balance_j\cdot\delta(\labeling_j,\labeling).
\end{eqnarray}
Preferences $\diffusion$ correspond to defensive label propagation~\cite{SB11d} that increases the propagation strength from the core of each group or, equivalently, decreases the strength of its border. This forces the algorithm to more gradually reveal the underlying structure and improves group detection in real-world networks~\cite{SB11d}. The core and the border of each group are modeled by employing random walks (see~\algref{gpa}).

Updates of nodes' labels in \equref{lpa} and \equref{dpa} occur sequentially, in a random order, to avoid label oscillations in, e.g., bipartite or star-like graphs. Still, this severely hampers the stability of the approach~\cite{RAK07,SB11b}, since nodes that are updated at the beginning of some iteration gain higher propagation strength than those that are considered last. 

Balancers $\balance$~\cite{SB11b} decrease (increase) the strength of the nodes that are considered first (last), which counteracts for the introduced randomness and stabilizes the algorithm. Let $\order_i\in(0,1]$ be a normalized index of node $\node_i$ in some random order. Assuming linearity, we can model node balancers simply as $\balance_i=\order_i$, however, the use of a sigmoid curve allows for some further control~\cite{SB11g} (see~\algref{gpa}, line~7). The latter in fact introduces parameters $\lambda$ and $\stability\geq0$, where $\lambda$ is fixed to $0.5$~\cite{SB11b}, while $\stability$ represents the balance between the algorithm stability and complexity (see~\secref{analysis}).

As label propagation in~\equref{dpa} can discover only densely connected groups of nodes, it is limited to communities. Nevertheless, the same principle can be extended to modules of nodes~\cite{SB12u}. Rather than propagating the labels between the neighboring nodes, labels are propagated between the nodes at distance two (i.e., through common neighbors). Since nodes in modules share many neighbors, similarly as before, they form a consensus on some particular label $\labeling$. Thus, when the propagation process unfolds, $\{\group\}$ contains modules that are well depicted in the structure of the network~\cite{SB12u,SB11g}.

\begin{eqnarray}
\label{equ:gpa}
\labeling_i & = & \argmax_\labeling \left( \parameter_\labeling \cdot \overbrace{\sum_{\node_j\in\neighbors_i} \balance_j\diffusion_j\cdot\delta(\labeling_j,\labeling)}^{\mbox{Community detection}} +\mbox{ }(1-\parameter_\labeling) \cdot \overbrace{\sum_{\substack{\node_j\in\neighbors_i\\ \node_k\in\neighbors_j\backslash\neighbors_i}} \frac{\balance_k\diffusiont_k}{\degree_j}\cdot\delta(\labeling_k,\labeling)}^{\mbox{Module detection}} \right)
\end{eqnarray}

The algorithm for detection of modules is shown in the right-hand side of~\equref{gpa}. Preferences $\diffusiont$ correspond to defensive propagation as above, while $\degree_j$ in denominator makes the sum proportional to $\degree_i$. Otherwise, \equref{gpa} represents a general group detection algorithm denoted General Propagation Algorithm~\cite{SB12u} (\GPA), where intrinsic parameters $\parameter\in [0,1]$ represent the adopted network modeling. Setting $\parameter_{\labeling}=1$ for all $\labeling$ is identical to community detection approach in~\equref{dpa}, whereas, for all $\parameter_{\labeling}$ equal to zero, the algorithm can reveal only modules. When $\parameter_{\labeling}=0.5$, identified groups are based on community and module links.

\begin{algorithm}[t]
\caption{\label{alg:gpa}General Propagation Algorithm (\GPA)}
\begin{algorithmic}[1]
\Require Graph $\graph(\nodes,\links)$ and parameters $\lambda$, $\stability$ ($\parameter$)
\Ensure Group partition $\{\group\}$ (i.e., labels $\labeling$)
\For{$\node_i\in\nodes$}
	\State $\labeling_i \gets i$ \Comment{Label initialization.}
	\State $\diffusion_i, \diffusiont_i \gets 1/\n$
\EndFor
\While{\textbf{not} \Call{Convergence}{}}
	\For{$\node_i\in\Call{Shuffle}{$\nodes$}$}
		\State $\balance_i \gets 1 / (1 + e^{-\stability (\order_i - \lambda)})$
		\State $\labeling_i \gets$ \equref{gpa} \Comment{Label propagation.}
		\State $\diffusion_i \gets \sum_{\node_j\in\neighbors_i} \diffusion_j \cdot \delta(\labeling_j,\labeling_i) / \sum_{\node_k\in\neighbors_j} \delta(\labeling_k,\labeling_i)$
		\State $\diffusiont_i \gets \sum_{\substack{\node_j\in\neighbors_i\\ \node_k\in\neighbors_j\backslash\neighbors_i}} \diffusiont_k \cdot \delta(\labeling_k,\labeling_i) / \sum_{\substack{\node_j\in\neighbors_k\\ \node_l\in\neighbors_j\backslash\neighbors_k}} \delta(\labeling_l,\labeling_i)$
	\EndFor
\EndWhile
\State \Return $\{\group_i\}$
\end{algorithmic}
\end{algorithm}

Group detection framework~\cite{SB12u,SB11g} in~\algref{gpa} is taken as the basis for the algorithm proposed in this paper. It ought, however, to be mentioned that one could else adopt an arbitrary group detection approach.

\subsection{Hierarchical Propagation Algorithm} \label{sec:algorithm:hpa}
Group detection framework in~\algref{gpa} can reliably discover communities and also modules, when they are well defined in the network structure (see~\secref{analysis}). Still, group parameters $\parameter$ that represent the adopted network modeling strategy have to be set accordingly. Approach in~\cite{SB12u} estimates values of $\parameter_{\labeling}$ during the propagation process by measuring the conductance~\cite{Bol98} of each group $\labeling$. Similarly, approach in~\cite{SB11g} is based on a the standard clustering coefficient $\clustering$~\cite{WS98}. However, due to an unsupervised nature of these strategies, both approaches suffer from the drifting effect, especially in larger networks.

We here propose a simpler strategy that fixes parameters $\parameter$ apriori during the label initialization step (\algref{gpa}, line~2). Recall that according to~\secref{background}, communities appear in dense parts of real-world networks, when the density is measured by the degree-corrected clustering coefficient $\dclustering$~\cite{SV05}. On the contrary, clear modules appear in sparse parts with lower $\dclustering$. Thus, when label $\labeling_i$ of node $\node_i$ is initialized, 
\begin{subnumcases}{\label{equ:model}\parameter_{\labeling_i} =}
1 & if $\dclustering_i\geq\random$ ($\DCLUSTERING\geq\random$), \label{equ:model:community} \\
0 & if $\dclustering_i<\random$ ($\DCLUSTERING<\random$), \label{equ:model:module} \\
0.5 & else, \label{equ:model:group}
\end{subnumcases}
where $\random$ is the expected network clustering in a random graph with the same degree sequence~\cite{NSW01}.\footnote{$\random$ is derived for an alternative, but similar, definition of network clustering~$\CLUSTERING$.}
\begin{eqnarray}
\label{equ:random}
\random & = & \frac{\left(\sum_i \degree_i^2 - \mean{\degree}\n \right)^2}{\mean{\degree}^3\n^3}
\end{eqnarray}

Since most real-world networks are dense and small-world with $\DCLUSTERING\gg\random$, \equref{model:community} reveals communities in dense regions of the networks. On the other hand, \equref{model:module} properly models, e.g., two-mode networks with $\DCLUSTERING=0$, where only modules are expected. However, as modules can also appear in denser regions with $\dclustering>\random$ (see~\secref{background:clustering}), the above strategy apparently ignores them. Nevertheless, the approach we propose below intentionally first discovers dependent modules (e.g., bipartite configurations) as merged into communities, when their presence is not immediately apparent from the structure of the network (with respect to $\dclustering$).

We introduce a group refinement procedure for the propagation framework in~\algref{gpa}.  Let $\{\group\}$ be an initial group partition revealed by the algorithm. Then, for each group $\group$, $|\group|\geq 3$, the algorithm is further applied to a network induced by the nodes in~$\group$ denoted $\graph(\group)$. As this refinements proceed recursively, an entire sub-hierarchy of groups is revealed for each group $\group$. Similarly, we reveal a super-hierarchy by applying the algorithm to a network induced by the initial groups $\{\group\}$. Here nodes in fact represent groups that are linked, when a link also exists in the original network. Final result of such hierarchical refinements and agglomerations is a complete hierarchy of groups $\hierarchy$ (see~\figref{AFL:hierarchy}). 

The presented approach reveals group hierarchy in a top-down and bottom-up fashion simultaneously. Whereas hierarchical investigation is somewhat common in community detection~\cite{GN02,CNM04}, top-down group refinement has not yet been used particularly for the discovery of modules (for top-down community discovery see, e.g.,~\cite{GN02,New06c}). However, since modules in real-world networks are often not immediately apparent from the network structure (see~\secref{analysis:realworld}), group refinement step is in fact crucial. An adequate group refinement procedure is thus, together with an improved network modeling strategy in~\equref{model}, the main novelty of the proposed approach denoted Hierarchical Propagation Algorithm (\HPA).

\begin{algorithm}[t]
\caption{\label{alg:hpa}Hierarchical Propagation Algorithm (\HPA)}
\begin{algorithmic}[1]
\Require Graph $\graph(\nodes,\links)$
\Ensure Group hierarchy $\hierarchy$
\State $\hierarchy \gets$ \Call{Hierarchy}{$\graph$} \Comment{Hierarchy discovery.}
\If{$|\graph(\hierarchy)|>1$}
	\State $\hierarchy \gets \hierarchy$ $\cup$ \Call{HPA}{$\graph(\hierarchy)$} \Comment{Group agglomeration.}
\EndIf
\State \Return $\hierarchy$
\Procedure{Hierarchy}{$\graph$}
	\State $\{\group_i\} \gets$\Call{Propagation}{$\graph$} \Comment{Group detection.}
	\If{$|\{\group_i\}|=1$}
		\State \Return $\hierarchy(\group)$
	\EndIf
	\State $\hierarchy \gets \{\}$
	\ForAll{$\{\group_i\}$}
		\State $\hierarchy_{\group_i} \gets$ \Call{Hierarchy}{$\graph(\group_i)$} \Comment{Group refinement.}
		\If{$\likelihood(\hierarchy_{\group_i})>\likelihood(\hierarchy(\group_i)) \wedge \graph(\group_i)\mbox{ connected}$}
			\State $\hierarchy \gets \hierarchy\cup\hierarchy_{\group_i}$
		\Else
			\State $\hierarchy \gets \hierarchy\cup\hierarchy(\group_i)$
		\EndIf
	\EndFor
	\State \Return $\hierarchy$
\EndProcedure
\end{algorithmic}
\end{algorithm}

Details of the algorithm are given in~\algref{hpa}. Observe that each revealed hierarchy $\hierarchy_{\group}$ is compared against an alternative with a single group $\group$ denoted $\hierarchy(\group)$ (\algref{hpa}, line~14). The hypothesis is tested using the likelihood $\likelihood\in(0,1]$ of the hierarchies given the network observed~\cite{CMN06}. Hence,
\begin{eqnarray}
\label{equ:likelihood}
\likelihood(\hierarchy) & = & \prod_{\inner_i\in\hierarchy} {\probability_i}^{\m_i} \cdot \left(1 - \probability_i\right)^{\M_i-\m_i},
\end{eqnarray}
where $\inner_i$ are inner nodes of $\hierarchy$ representing different groups of nodes $\group_i$, $\m_i$ is the number of links between the groups represented by the sub-hierarchy rooted at $\inner_i$, $\M_i$ is the number of all possible links and $\probability_i=\m_i/\M_i$ are the maximum likelihood estimators for $\hierarchy$~\cite{CB90}. \LOGL of a hierarchy with a single level is in fact the entropy of a corresponding blockmodel~\cite{Pei12c}, while in~\secref{comparison:realworld} we report \MLOGL, where smaller is better.

\HPA algorithm reveals an entire group hierarchy $\hierarchy$, thus, when only a partition is required, we report the groups represented by the bottom-most inner nodes in $\hierarchy$ (no group agglomerations are needed).


\section{Analysis of the algorithm} \label{sec:analysis}
The proposed \HPA algorithm is rigorously analyzed and compared against different alternatives.

\subsection{Complexity, stability and validity} \label{sec:analysis:complexity}
We first analyze the behavior of the proposed algorithm with respect to the only parameter $\stability$ corresponding to balanced propagation (see~\secref{algorithm:gpa}). \tblref{analysis:realworld} shows pairwise Normalized Variation of Information~\cite{KLN08} (\NVI) of the revealed group partitions for two real-world networks (see~\tblref{realworld}). \NVI is a distance in the space of partitions, while lower values represent better correlation, \NVI$\in[0,1]$. Besides \HPA algorithm, we also include the basic label propagation in~\equref{lpa} (\LPA algorithm) and a general group detection approach in~\equref{gpa} (\GPA algorithm). Note that \HPA and \GPA algorithms differ merely in the hierarchical group refinement procedure introduced in~\secref{algorithm:hpa}.

\begin{table}[htb]
\centering
\caption{\label{tbl:analysis:realworld}Mean pairwise \NVI of the revealed group partitions~(and the number of algorithm iterations) over $100$~runs.}
\begin{tabular}{ccccc} \HLINE
Network & Algorithm & $\stability=0$ & $\stability=0.25$ & $\stability=2$ \\\TLINE
\multirow{3}{*}{\AFL} & \LPA & {$0.099$ ($3.6$)} & {\NULL} & {\NULL} \\
 & \GPA & {$0.090$ ($3.6$)} & {$0.078$ ($3.9$)} & {$\mathbf{0.060}$ ($7.1$)} \\
 & \HPA & {$0.082$ ($5.6$)} & {$0.073$ ($5.8$)} & {$\mathbf{0.065}$ ($9.2$)} \\\HLINE
\multirow{3}{*}{\SWC} & \LPA & {$0.260$ ($4.3$)} & {\NULL} & {\NULL} \\
 & \GPA & {$0.092$ ($3.1$)} &{$0.086$ ($3.2$)} & {$\mathbf{0.064}$ ($3.8$)} \\
 & \HPA &{$0.094$ ($3.1$)} & {$0.083$ ($3.2$)} & {$\mathbf{0.061}$ ($4.5$)} \\\HLINE
\end{tabular}
\end{table}

While setting $\stability=0$ is identical to the standard propagation, increasing parameter $\stability$ improves the stability of all algorithms and, consequently, also the robustness of the revealed group structure (see \tblref{analysis:realworld}). However, the number of iterations needed for the propagation to converge also increases. Parameter $\stability$ thus controls the balance between the algorithm's stability and complexity. For the comparison in \secref{comparison} we set $\stability=2$, and $\stability=0.25$ for larger real-world networks, while $\stability=0$ for the analysis below.

We further analyze the complexity of the proposed approach. Each iteration of \LPA algorithm for community detection takes $\complexity{\m}$, while the number of iterations grows very slowly as, e.g., $\complexity{\log\m}$~\cite{LHLC09}. Algorithm complexity was also estimated to $\complexity{\m^{1.23}}$~\cite{SB11d}. On the other hand, the complexity of each iteration of the general group detection algorithms \GPA and \HPA is obviously $\complexity{\mean{\degree}\m}$. 

\begin{figure}[tb]
\centering
\subfloat[\label{fig:models:FF}Graph model]{
\includegraphics[width=\plotwidth]{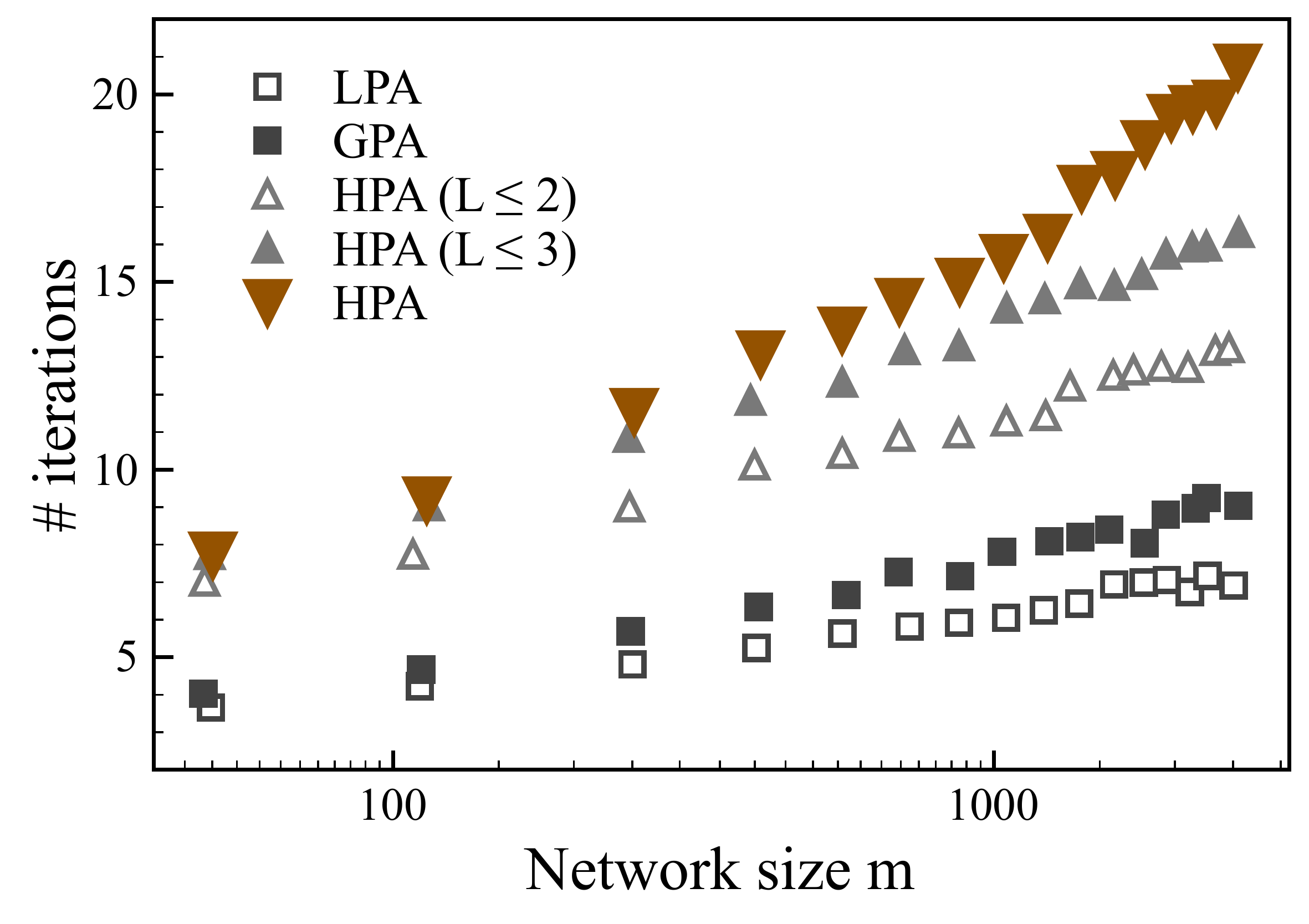}}
\subfloat[\label{fig:models:ER}Random graph]{
\includegraphics[width=\plotwidth]{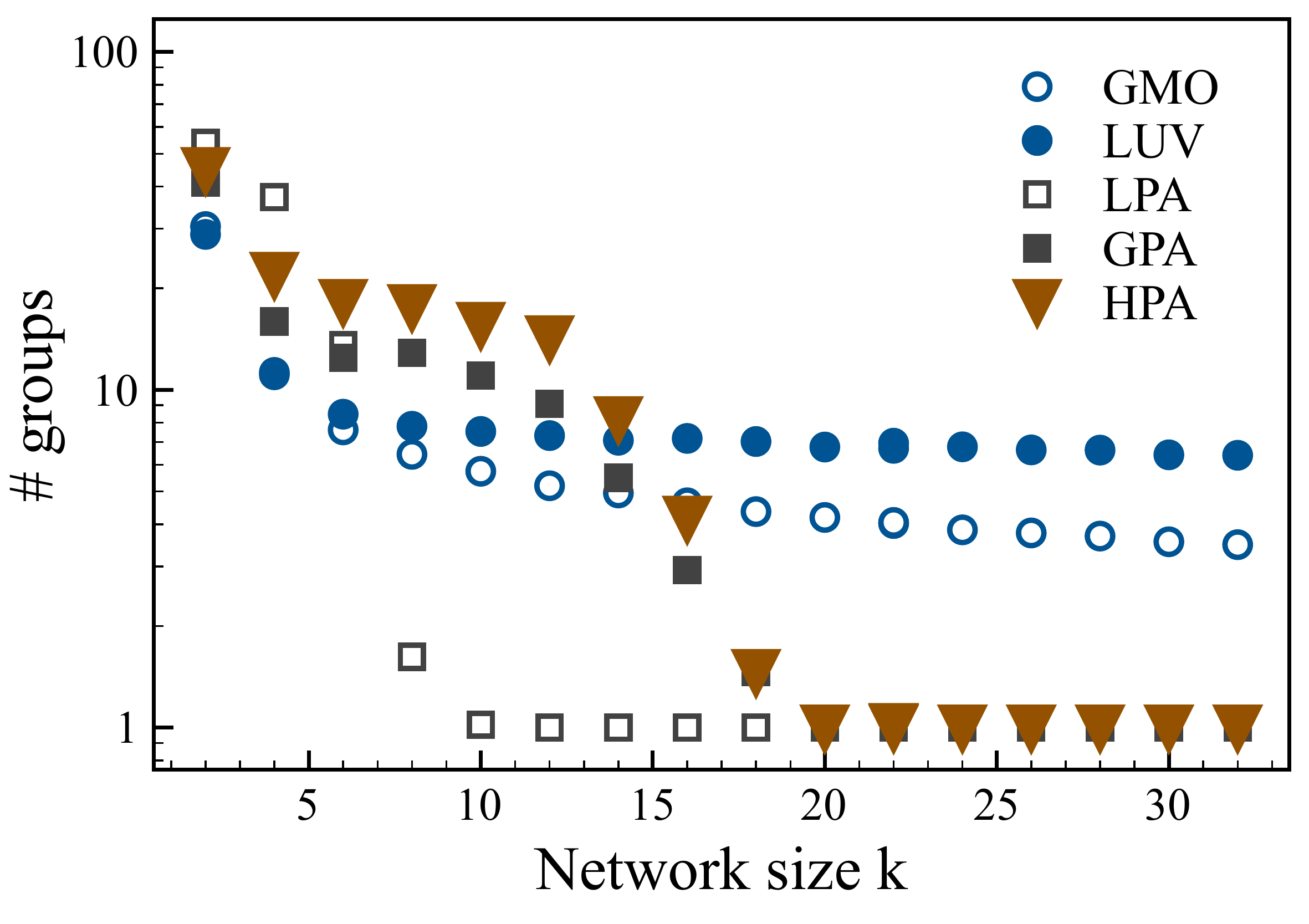}}
\caption{\label{fig:models}Mean number of (a)~algorithm iterations and (b)~revealed groups over $100$ graph realizations.}
\end{figure}

\figref{models:FF} shows the number of iterations for different algorithms applied to a realistic graph model~\cite{LKF07}, where the parameter $\levels$ in the legend refers to the number of nontrivial levels in the revealed group hierarchy. Although general group detection and hierarchical investigation indeed increase complexity of the propagation, the number of iterations still appears to grow no faster than $\complexity{\log\m}$ (mind the scales). There is a super-logarithmic increase in the case of \HPA algorithm, but analysis on real-world networks in \secref{comparison:realworld} reveals that $\levels$ is commonly only slightly larger than in the case of a comparable community detection approach (i.e., only one or two refinements steps are needed). Thus, the complexity of the proposed \HPA algorithm can be estimated to $\complexity{\mean{\degree}\m\log\m}$, which should scale to networks with a million links.

Last, we also validate \HPA~algorithm on Erd\"{o}s-R\'{e}nyi~\cite{ER59} random graphs that presumably contain no characteristic groups of nodes (\figref{models:ER}). When the network size exceeds a certain threshold, the algorithm reveals no group structure (i.e., classifies all nodes into a single group). Behavior of the general group detection approaches else differs from \LPA algorithm for communities only, nevertheless, the proposed approach does not suffer from the same problems like, e.g., modularity~\cite{NG04} optimization~\cite{FB07} exemplified by \GMO and \LUV algorithms (see~\secref{comparison:framework}).

\subsection{Hierarchical group detection} \label{sec:analysis:algorithm}
\begin{figure}[tb]
\centering
\subfloat[\label{fig:analysis:GN}\GN benchmark]{
\includegraphics[width=\plotwidth]{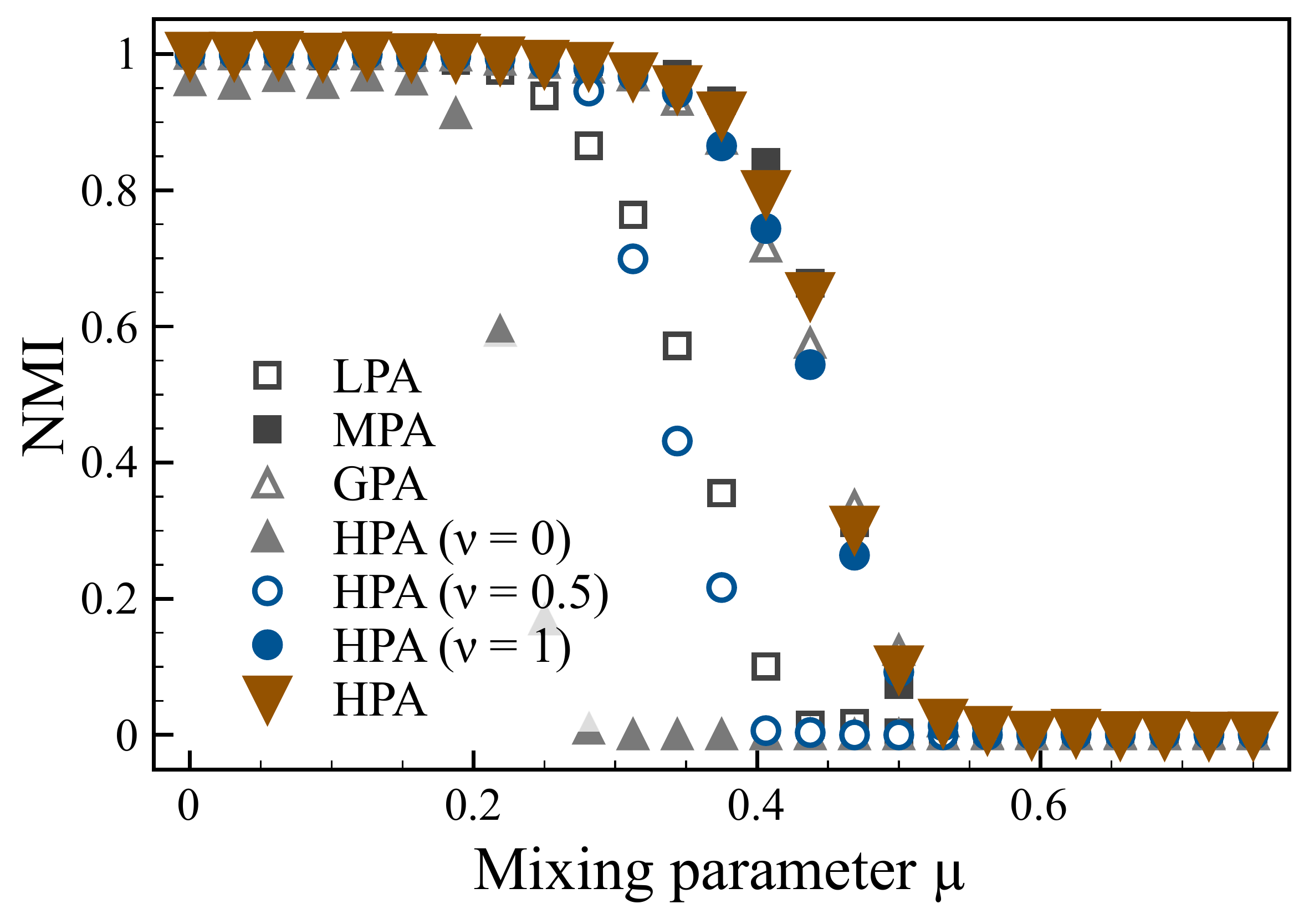}}
\subfloat[\label{fig:analysis:GN2}\GNT benchmark]{
\includegraphics[width=\plotwidth]{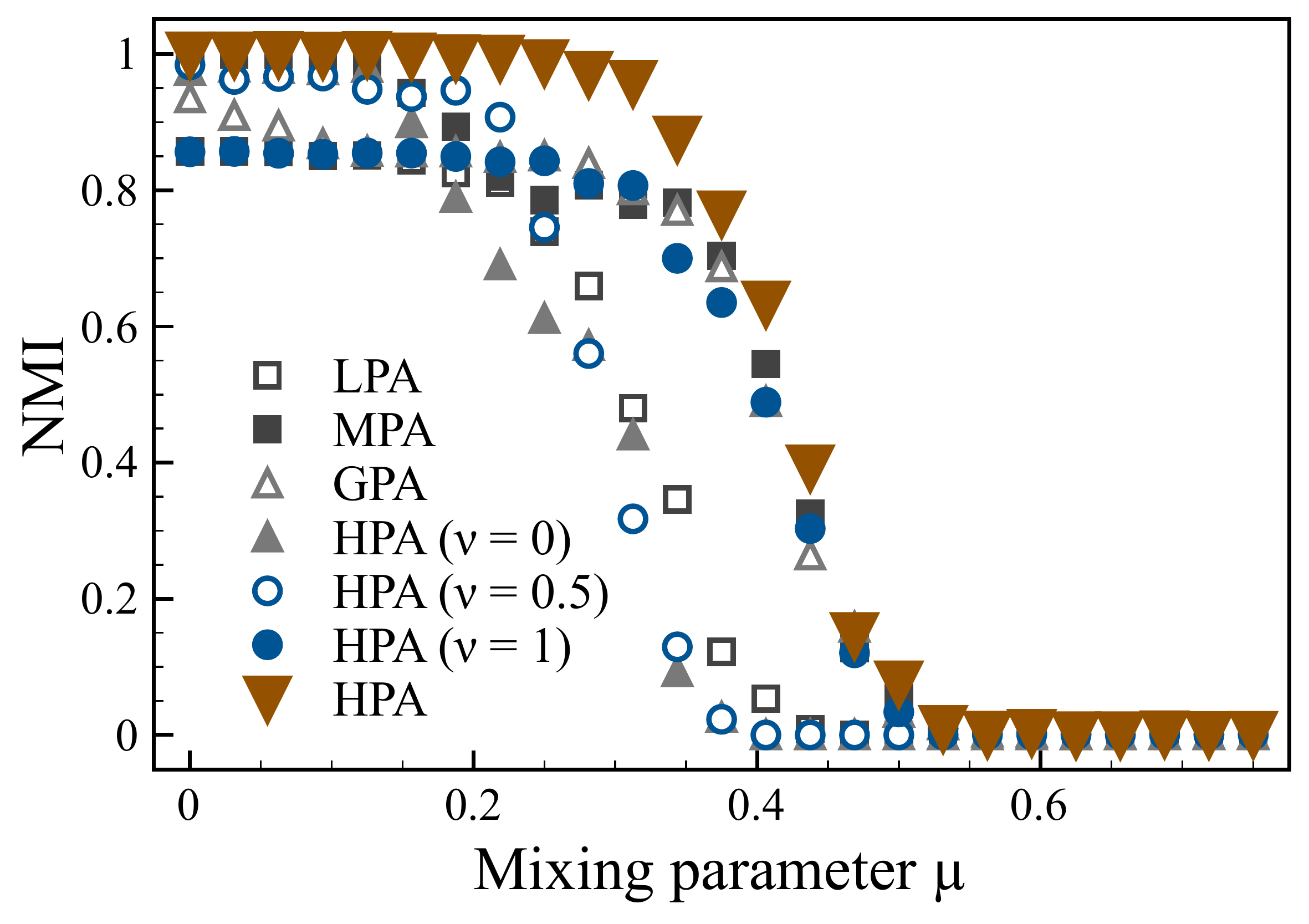}}
\caption{\label{fig:analysis}Mean \NMI for group detection task over $100$ graph realizations.}
\end{figure}

Below we compare \HPA algorithm against several alternatives on two synthetic benchmark graphs (see~\tblref{synthetic}) with planted partitions of four communities (\figref{analysis:GN}) and two communities and two modules (\figref{analysis:GN2}). Distribution of links is controlled by a mixing parameter $\mixing\in[0,1]$, where lower values correspond to a clearer group structure. 
Results are reported in the form of Normalized Mutual Information~\cite{DDDA05} (\NMI), which has become a \textit{de facto} standard in the literature, \NMI$\in[0,1]$ (higher is better).

Observe that \HPA algorithm limited to module detection (i.e., $\parameter=0$) cannot accurately reveal communities for larger $\mixing$ (see~\figref{analysis:GN}), while community detection approach (i.e., $\parameter=1$) completely fails at module detection (see~\figref{analysis:GN2}). On the other hand, approach with $\parameter=0.5$ can detect general groups of nodes but only for smaller $\mixing$. Only the proposed \HPA algorithm can accurately reveal both communities and modules, and clearly outperforms all other alternatives including the best existing network group modeling strategy~\cite{SB11g} (i.e, \MPA algorithm). Note also the difference between proposed approach and \GPA algorithm in~\figref{analysis:GN2}, which corresponds to the increase of performance due to an adequate hierarchical group refinement. Particularly, both approaches detect three dense groups of nodes (i.e., communities), whereas only \HPA algorithm further refines one of these into two modules that were in fact planted into the network.

\subsection{Football social network} \label{sec:analysis:realworld}
\begin{figure}[tb]
\centering
\subfloat[\label{fig:AFL:network}\AFL network]{
\includegraphics[width=\smallwidth]{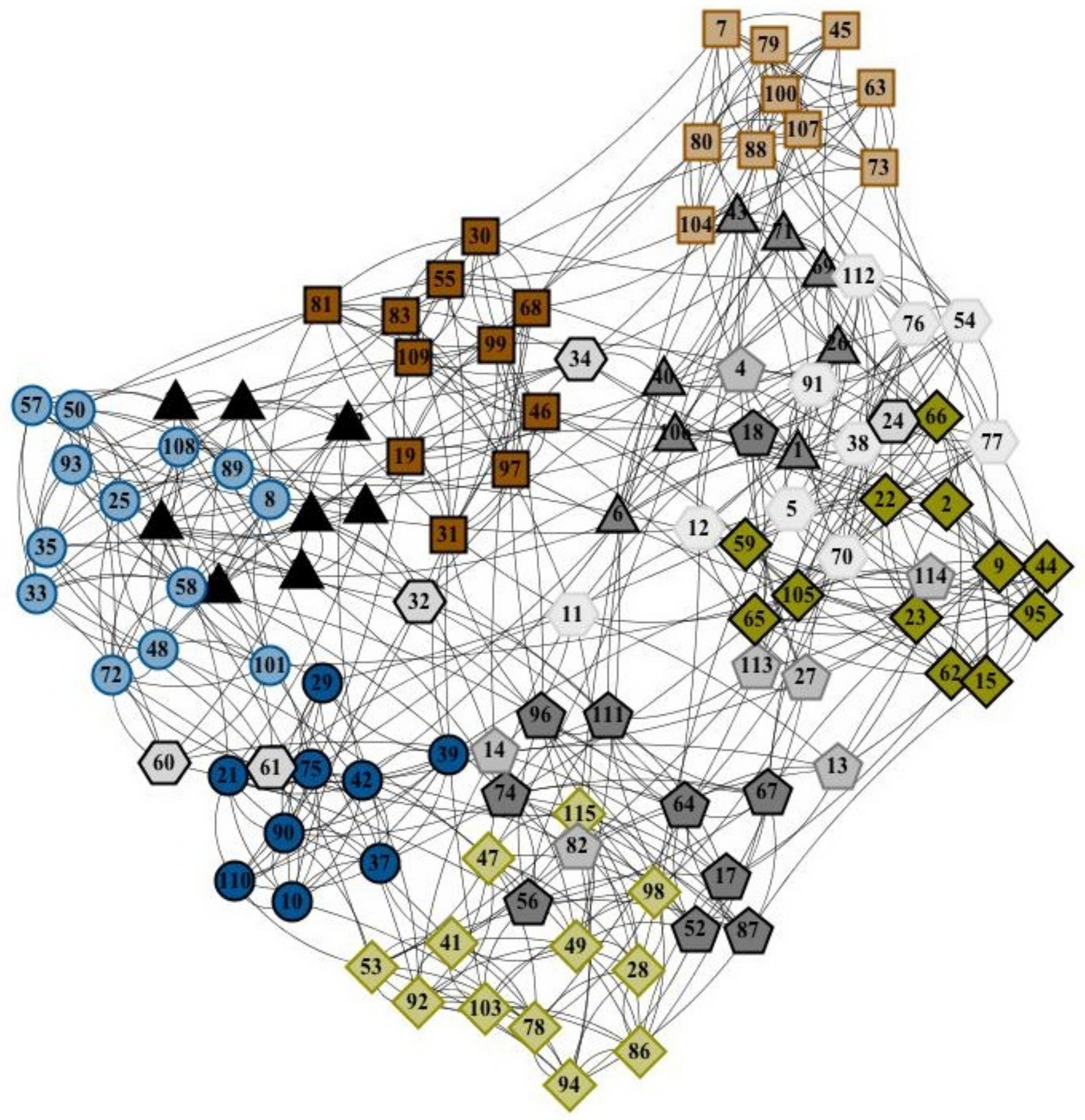}}
\subfloat[\label{fig:AFL:hierarchy}Revealed with \HPA]{
\includegraphics[width=\figurewidth]{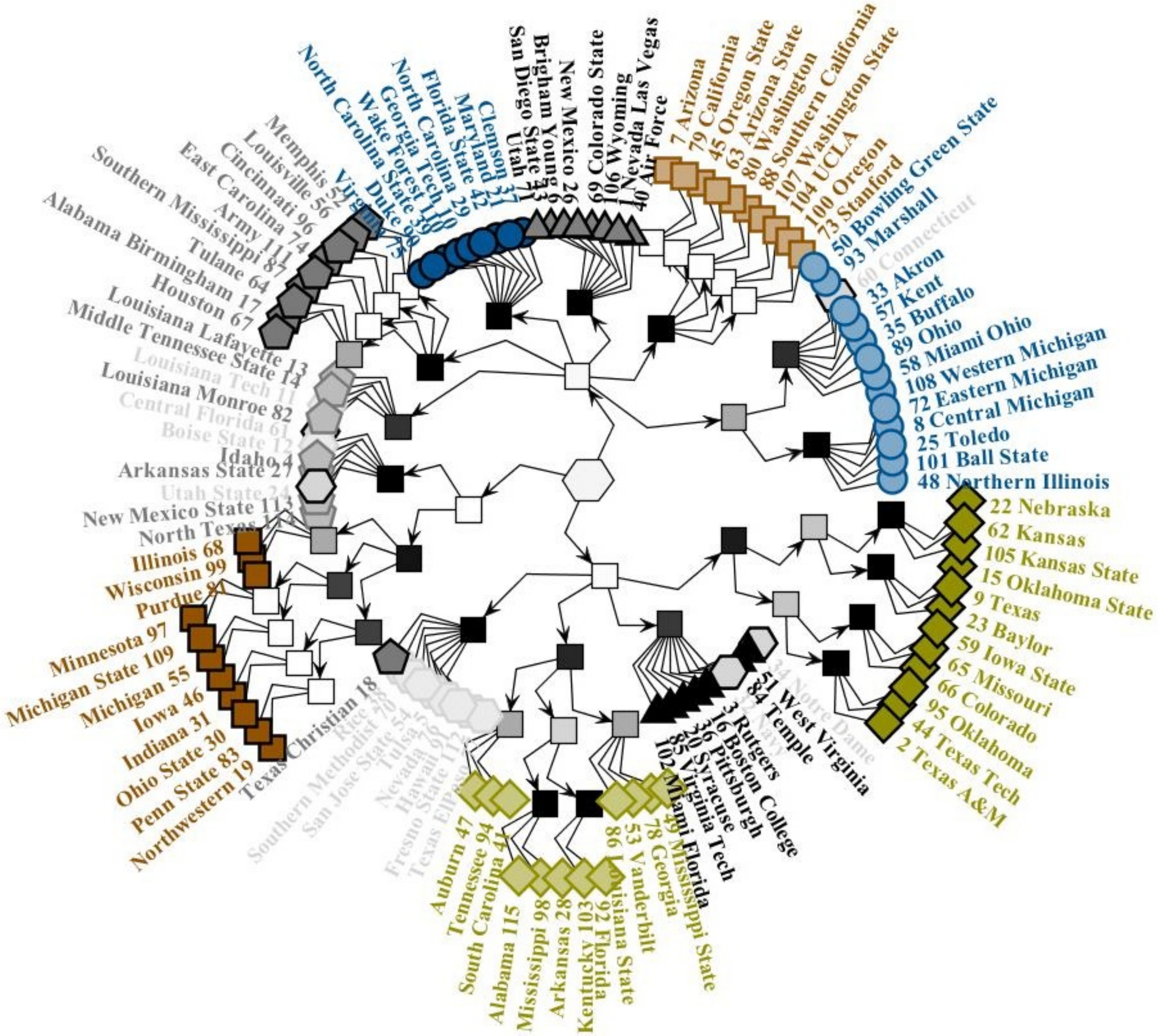}}
\caption{\label{fig:AFL}Group hierarchy of \AFL network with \logl{-970.2}, where node shapes correspond to a known sociological division into groups. (Shades of the inner nodes of the hierarchy are proportional to probabilities $\probability$.)}
\end{figure}

\figref{AFL} further demonstrates the benefits of a general group detection approach like \HPA algorithm. We apply the algorithm to a famous \AFL social network that represents matches played among US college football teams in the $2000$ season~\cite{GN02} (see~\tblref{realworld}). The latter has been of considerable interest in the community detection literature in the past, since the network reveals clear communities that coincide with the division into conferences. Revealed group hierarchy in~\figref{AFL:hierarchy} indeed contains communities on higher levels, however, several of these are then refined into well defined configurations of modules on lower levels. For instance, group of nodes at the top of~\figref{AFL:network} is in fact a complete multi-partite graph on ten nodes. Whether the particular group hierarchy would be of any interest in practical applications remains unclear, still, most of the groups present would remain overlooked under the standard community framework.

\section{Comparison of the state-of-the-art} \label{sec:comparison}
The proposed \HPA algorithm is compared against the current state-of-the-art approaches in network group detection. In the following, we first present the experimental framework.

\subsection{Experimental framework} \label{sec:comparison:framework}
Different approaches are first compared on synthetic benchmark graphs with planted group partition (\tblref{synthetic}). We adopt classical community detection benchmark that contains four equally-sized communities~\cite{GN02} (\GN benchmark) and two variations of a more realistic benchmark graphs with scale-free degree and community size distributions~\cite{LFR08} (\LFR benchmarks). For a general group detection task, we adopt a generalization of the \GN benchmark with two communities and two modules that are combined into a bipartite configuration~\cite{SB12u} (\GNT benchmark), and also two benchmarks with heterogeneous group size distribution and planted communities and bipartite or tripartite configurations (\SBV and \SBX benchmarks).

\begin{table}[htb]
\centering
\caption{\label{tbl:synthetic}Synthetic benchmark graphs used in the comparison of the state-of-the-art (``Groups'' corresponds to either the number of communities and modules or community~sizes).}
\begin{tabular}{ccccc} \HLINE
Graph & Description & $\n$ & $\approx\m$ & Groups \\\TLINE
\GN & {Classical community benchmark~\cite{GN02}} & {$128$} & {$1024$} & {$4$ -- $0$} \\
\multirow{2}{*}{\LFR} & \multirow{2}{*}{\parbox{0.4\linewidth}{\centering Reliable community benchmark with power-law group size distribution~\cite{LFR08}}} & \multirow{2}{*}{$1000$} & \multirow{2}{*}{$9800$} & {$[10,50]$} \\
 & & & & {$[20,100]$} \\
\GNT& {Generalization of \GN benchmark~\cite{SB12u}} & {$128$} & {$1024$} & {$2$ -- $2$} \\
\SBV & {Partly bipartite graph with communities} & {$112$} & {$936$} & {$3$ -- $4$} \\
\SBX & {Partly tripartite graph with communities} & {$112$} & {$1448$} & {$3$ -- $3$} \\\HLINE
\end{tabular}
\end{table}

Besides synthetic benchmark graphs, we also compare the approaches on ten real-world networks (\tblref{realworld}). These were selected thus to include most types of networks that are commonly analyzed in the literature (e.g., social or information networks), whereas detailed description is omitted here. Note that, due to a large number of experimental settings in the paper, the networks are only of moderate size.

\begin{table}[htb]
\centering
\caption{\label{tbl:realworld}Real-world networks used in the comparison of the state-of-the-art (\SWC and \SCI are two-mode networks).}
\begin{tabular}{cccc} \HLINE
Network & Description & $\n$ & $\m$ \\\TLINE
\AFL & {US college football league~\cite{GN02}} & {$115$} & {$613$} \\
\NSC & {Network science co-authorships~\cite{New06b}} & {$1589$} & {$2742$} \\
\PLB & {Co-purchased political books~\cite{Kre06}} & {$105$} & {$441$} \\
\EUR & {European highway network~\cite{SB11b}} & {$1039$} & {$1305$} \\
\JNG & {JUNG network analysis library~\cite{SB11s}} & {$317$} & {$719$} \\
\CLT & {Colt scientific computing library~\cite{SB11s}} & {$520$} & {$3691$} \\
\JVX & {\texttt{javax} namespace of Java library~\cite{SB11s}} & {$1595$} & {$5287$} \\
\MNE & {C. elegans metabolic pathways~\cite{JTAOB00}} & {$453$} & {$2025$} \\
\SWC & {Southern women affiliations~\cite{DGG41}} & {$32$} & {$89$} \\
\SCI & {Scottish corporate interlocks~\cite{SH80}} & {$217$} & {$348$} \\\HLINE
\end{tabular}
\end{table}

\HPA algorithm is compared against twelve other state-of-the-art approaches for group detection. For community detection detection task, we consider the following: greedy optimization of modularity~\cite{CNM04} (\GMO algorithm), multi-stage modularity optimization or Louvain method~\cite{BGLL08} (\LUV algorithm), sequential clique percolation approach~\cite{KKKS08} (\SCP  algorithm), Markov clustering algorithm~\cite{Don00} (\MCL algorithm), structural compression approach or Infomod~\cite{RB07} (\IMD algorithm), random walk-based compression known as Infomap~\cite{RB08} (\IMP algorithm) and the basic label propagation algorithm~\cite{RAK07} (\LPA algorithm). (\SCP algorithm reveals overlapping groups, thus, each node in multiple groups is classified into a random one.)

For general group detection task, we adopt the following approaches: symmetric nonnegative matrix factorization~\cite{DLPP06} (\NMF algorithm), $k$-means data clustering~\cite{Mac67} based on~\cite{LKC10} (\KMN algorithm), mixture models (i.e., stochastic blockmodel) estimated by expectation-maximization algorithm~\cite{NL07} (\EMM algorithm), mixture models with degree corrections~\cite{KN11a} (\DMM algorithm), model-based propagation algorithm~\cite{SB11g} (\MPA algorithm), structural compression approach~\cite{RB07} (\IMD algorithm) and the best community detection algorithm above~\cite{RB08} (\IMP algorithm). Note that \NMF, \KMN, \EMM and \DMM algorithms demand the number of groups apriori, which is a big disadvantage in practice. (Due to stability issues, \NMF and \DMM algorithms are applied to each network ten times and the best revealed group partition is reported.)

Due to limited resources, we do not include some otherwise very prominent approaches like~\cite{HEPF10,NM11a}.

\subsection{Comparison on synthetic graphs} \label{sec:comparison:synthetic}
\begin{figure}[tb]
\centering
\subfloat[\label{fig:synthetic:GN}\GN benchmark]{
\includegraphics[width=\plotwidth]{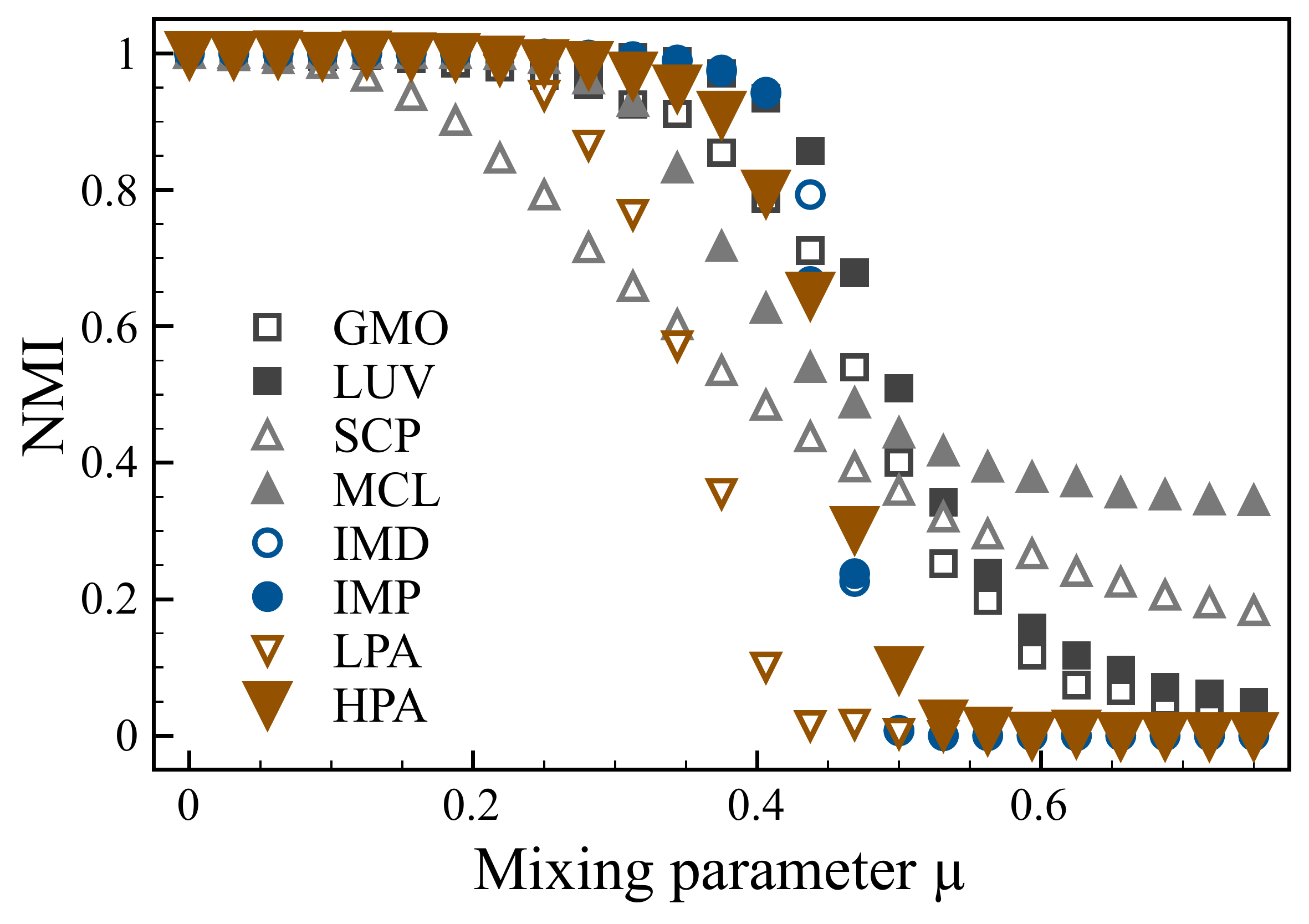}}
\subfloat[\label{fig:synthetic:LFRs}\LFR benchmark (small)]{
\includegraphics[width=\plotwidth]{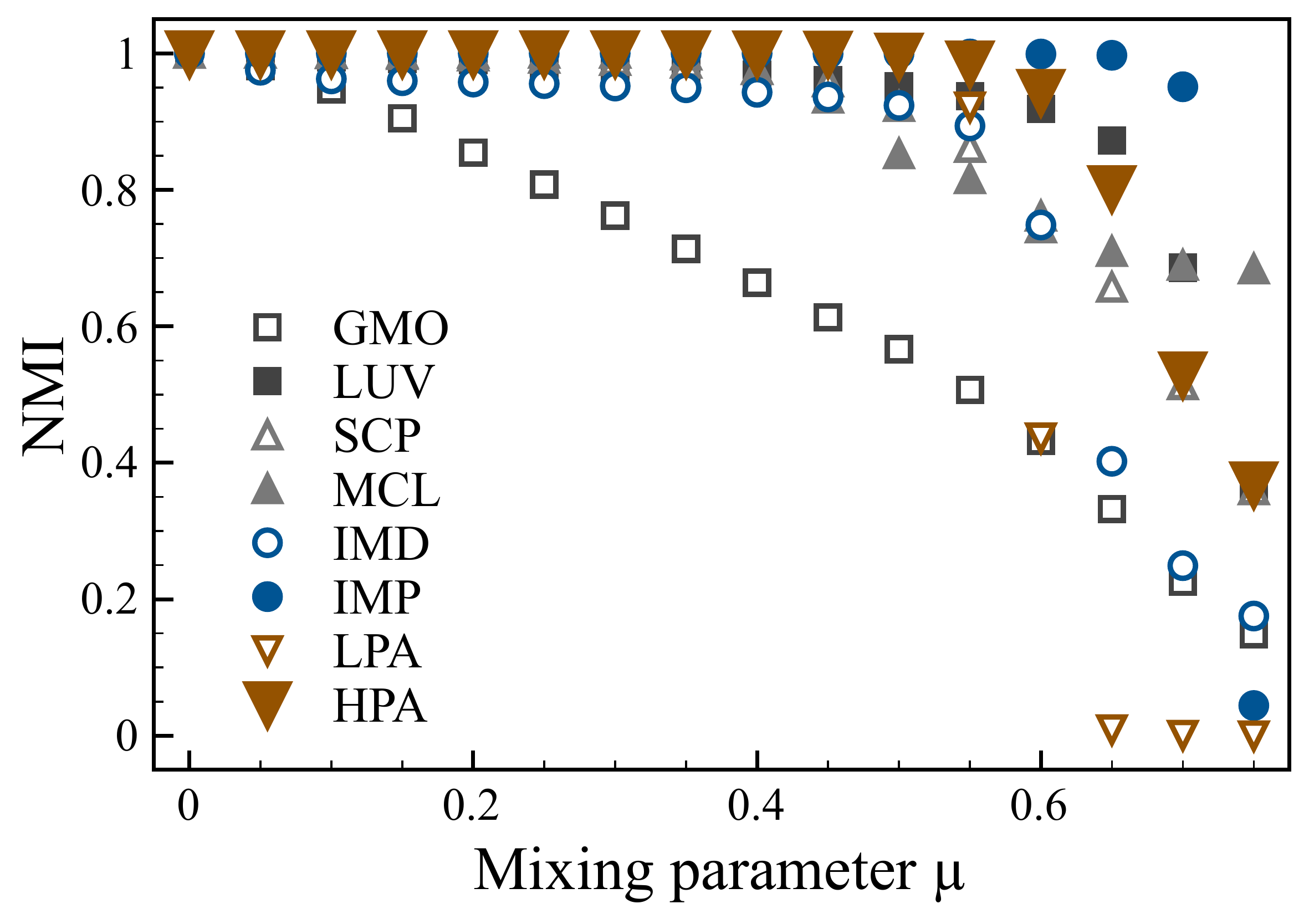}}
\subfloat[\label{fig:synthetic:LFRb}\LFR benchmark (big)]{
\includegraphics[width=\plotwidth]{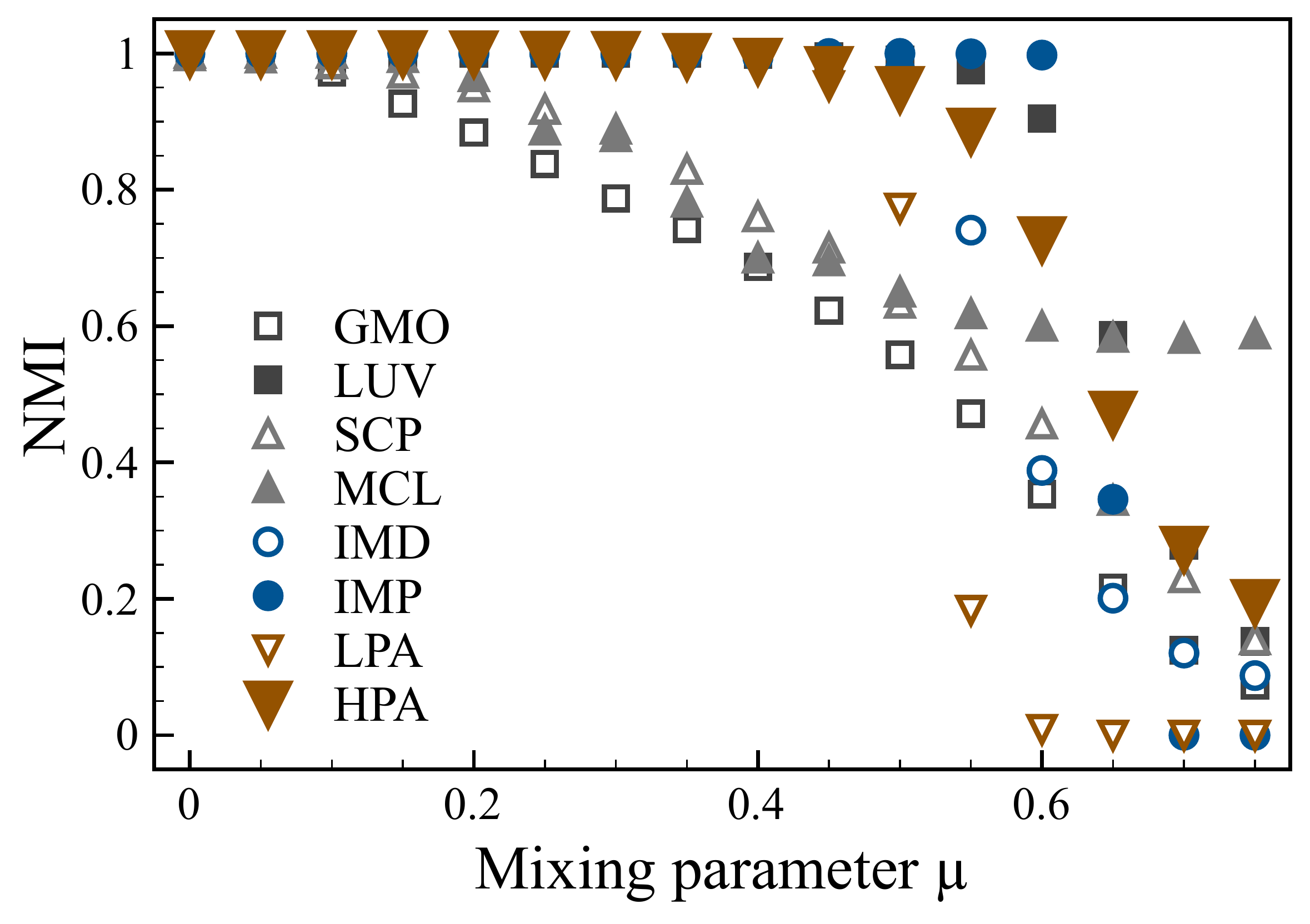}}\\
\subfloat[\label{fig:synthetic:GN2}\GNT benchmark]{
\includegraphics[width=\plotwidth]{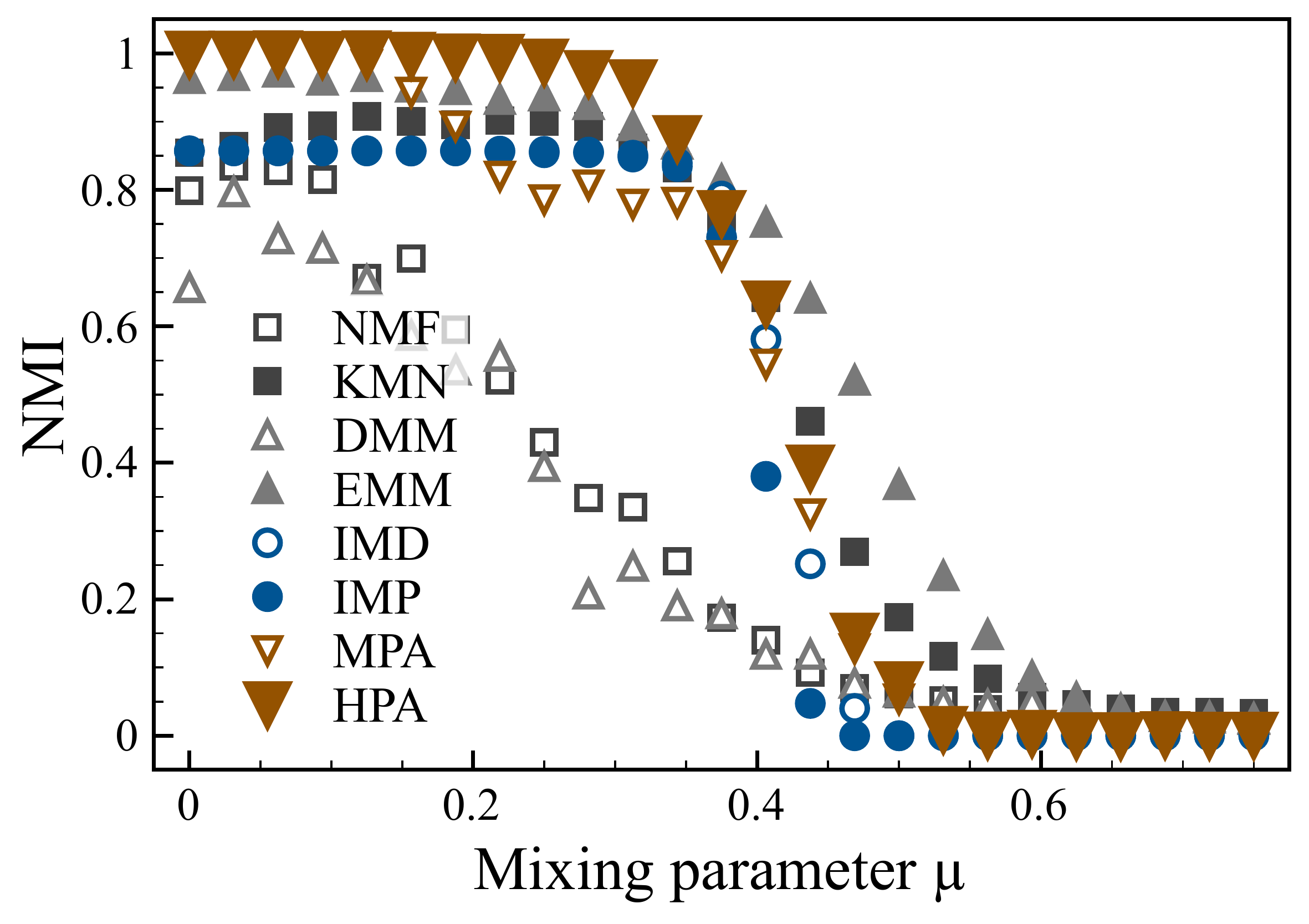}}
\subfloat[\label{fig:synthetic:SB7}\SBV benchmark]{
\includegraphics[width=\plotwidth]{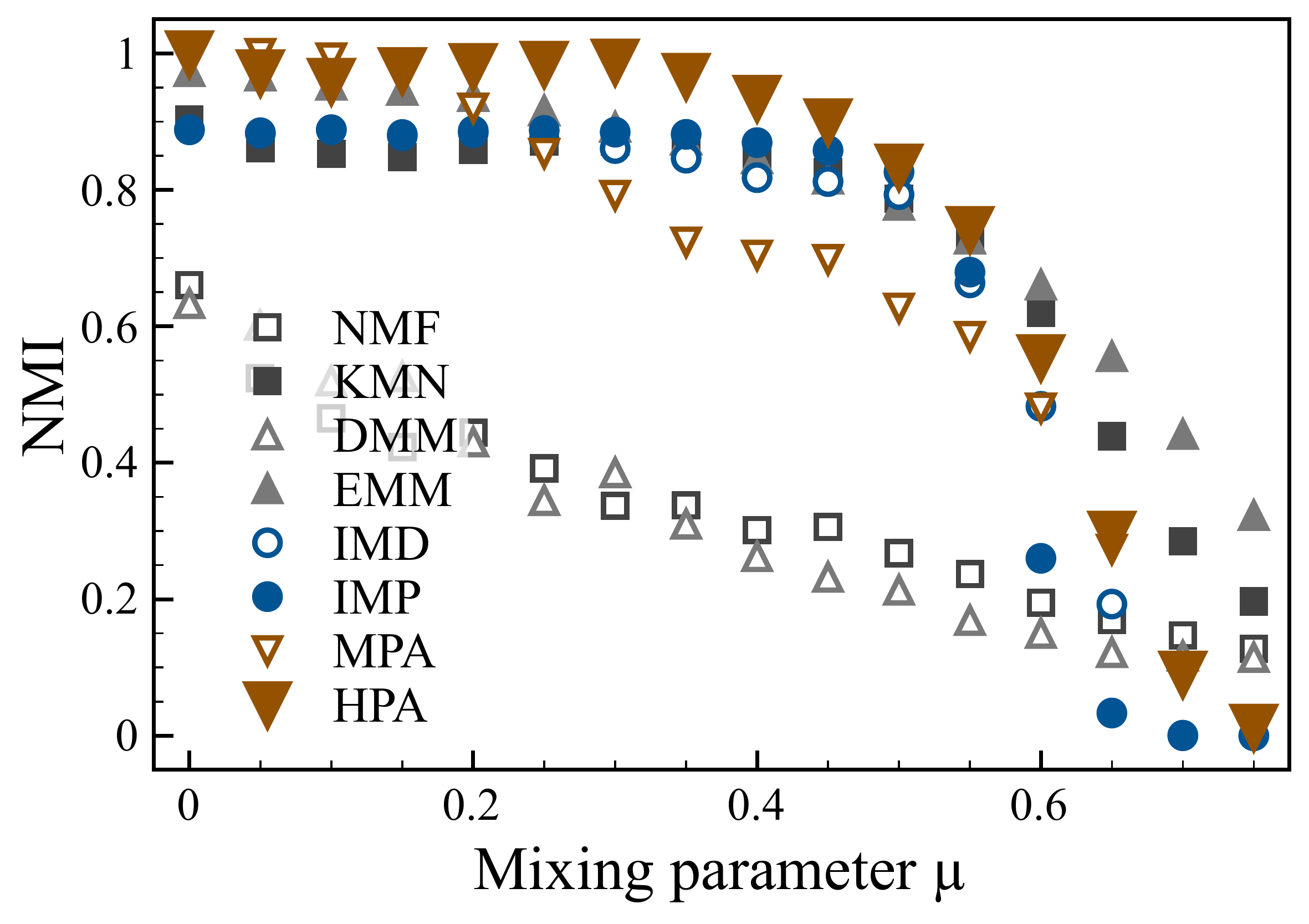}}
\subfloat[\label{fig:synthetic:SB6}\SBX benchmark]{
\includegraphics[width=\plotwidth]{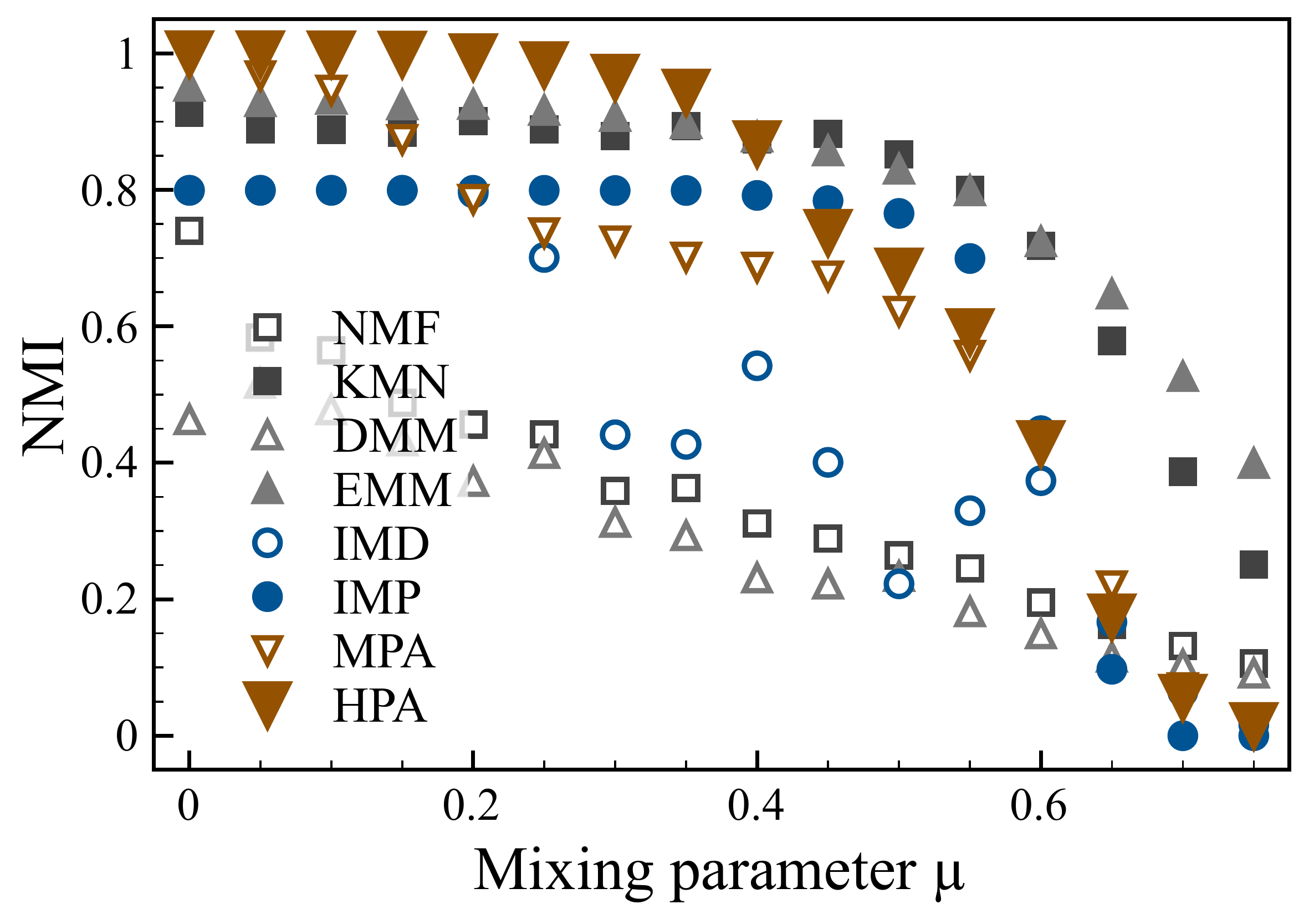}}
\caption{\label{fig:synthetic}Mean \NMI for group detection task over $100$ graph realizations ($10$ for \MCL and~\DMM algorithms).}
\end{figure}

\figref{synthetic} presents the results of the comparison on group detection task for the case of various synthetic benchmark graphs with planted group partition (see~\tblref{synthetic}). Benchmark graphs in the top row contain only communities, whereas those in bottom row contain also different configurations of modules. Group structure is again controlled by a mixing parameter $\mixing$, while results are reported as \NMI (\secref{analysis:algorithm}).

\IMP algorithm outperforms all other approaches in community detection (\figref{synthetic}, top), while the proposed \HPA algorithm is comparable to the second-best \LUV approach. Results are consistent with an empirical comparison of a larger number of approaches in~\cite{LF09b}, where none could improve on \IMP algorithm (\LUV algorithm performed similarly as above). \HPA algorithm can thus be considered at least comparable to the current state-of-the-art on community detection task.

Note, however, that none of the above community-specia\-lized approaches can discover general groups of nodes. In fact, only the proposed \HPA algorithm can accurately detect the planted modules in the bottom row of~\figref{synthetic} and notably improves on the existing \MPA algorithm. Although not immediately evident, data clustering approach (i.e., \KMN algorithm) performs surprisingly well on module detection task, but it fails in community detection, due to a loss of information about the network structure during transformation. Otherwise, only mixture models (i.e., \EMM algorithm) that were very popular in the literature lately can also detect different groups of nodes planted in these benchmark graphs. Still, observe that the performance significantly decreases in the case of groups with heterogeneous size distribution (e.g., \figref{synthetic:SB6}), whereas the improved model that corrects for node degrees (i.e., \DMM algorithm) fails on these benchmarks, due to increased complexity. Other approaches also do not perform well (e.g., \NMF algorithm). It should be stressed that despite relatively good performance of \KMN and \EMM algorithms, both approaches require the number of groups apriori, which seriously limits their applicability in practice~\cite{KN11a}. The latter does not hold for the proposed \HPA algorithm.

\subsection{Comparison on real-world networks} \label{sec:comparison:realworld}
In the following, we present results of the comparison of the state-of-the-art approaches on real-world networks. We consider network group detection, hierarchy discovery and link prediction tasks.

\subsubsection{Network group detection}
The approaches are first compared on group detection task on two real-world networks with a known sociological division into groups, i.e., \AFL social network with twelve communities introduced in \secref{analysis:realworld} and a two-mode \SWC affiliation network with three modules~\cite{DGG41} (see~\tblref{realworld}). Results are reported in the form of \NMI (\secref{analysis:algorithm}) and Adjusted Rand Index~\cite{HA85} (\ARI), where higher is better, \ARI$\in[0,1]$.

\begin{table}[htb]
\centering
\caption{\label{tbl:realworld:partition}Mean \NMI and \ARI for sociological group detection task over $100$ runs.}
\begin{tabular}{ccccccc} \HLINE
Network & \LPA & \LUV & \IMP & \KMN & \EMM & \HPA \\\TLINE
\multirow{2}{*}{\AFL} & {$0.892$} & {$0.876$} & {$\mathbf{0.922}$} & {$0.845$} & {$0.823$} & {$0.909$} \\
 & {$0.796$} & {$0.771$} & {$\mathbf{0.890}$} & {$0.698$} & {$0.683$} & {$0.850$} \\\HLINE
\multirow{2}{*}{\SWC} & {$0.184$} & {$0.309$} & {$0.417$} & {$0.677$} & {$0.827$} & {$\mathbf{0.932}$} \\
 & {$0.093$} & {$0.174$} & {$0.273$} & {$0.560$} & {$0.720$} & {$\mathbf{0.936}$} \\\HLINE
\end{tabular}
\end{table}

Results in~\tblref{realworld:partition} show similar performance as in the case of synthetic graphs in~\secref{comparison:synthetic}. \IMP algorithm is the best approach for community detection (i.e., \AFL network), while the proposed \HPA algorithm again performs well. In contrast to before, \HPA algorithm clearly outperforms \LUV algorithm (according to \ARI). Furthermore, \HPA algorithm is also a clear winner in module detection (i.e., \SWC network), whereas \KMN and \EMM algorithms fail on these networks.

\subsubsection{Network hierarchy discovery}
We further compare the entire hierarchies of groups revealed with different state-of-the-art approaches. For a fair comparison, all approaches are extended with the hierarchical group agglomeration procedure  proposed in \HPA algorithm. \tblref{realworld:likelihood} shows \LOGL-s that correspond to posterior probabilities of the revealed hierarchies, where higher values are better, \LOGL$\leq 0$ (\secref{algorithm:hpa}). Expectedly, group hierarchies discovered with \HPA algorithm appear most likely, still, this is at least partly due to a larger search space. However, as claimed in~\secref{analysis:complexity}, the number of nontrivial levels in the group hierarchies revealed with \HPA algorithm is only slightly larger than for a comparable community detection approach (e.g., \LUV algorithm).

\begin{table}[htb]
\centering
\caption{\label{tbl:realworld:likelihood}Peak \MLOGL (and the number of nontrivial levels) of the group hierarchies revealed over $100$ runs.}
\begin{tabular}{cccccc} \HLINE
Network & \LUV & \IMP & \EMM & \HPA \\\TLINE
\AFL & {$1119.3$ ($2$)} & {$1066.7$ ($1$)} &  {$1144.7$ ($2$)} & {$\mathbf{1016.8}$ ($3$)} \\
\NSC & {$2542.6$ ($3$)} & {$2106.3$ ($3$)} & {\NULL} & {$\mathbf{2041.4}$ ($4$)} \\
\EUR & {$5819.3$ ($2$)} & {$4410.2$ ($3$)} & {\NULL} & {$\mathbf{3981.0}$ ($5$)} \\
\JNG & {$2540.3$ ($2$)} & {$2302.9$ ($2$)} & {\NULL} & {$\mathbf{2259.9}$ ($4$)} \\
\JVX & {$13515.4$ ($2$)} & {$12438.3$ ($2$)} & {\NULL} & {$\mathbf{11512.2}$ ($3$)} \\
\SWC & {$196.4$ ($2$)} & {$203.2$ ($1$)} & {$\mathbf{163.6}$ ($1$)} & {$\mathbf{163.6}$ ($1$)} \\\HLINE
\end{tabular}
\end{table}

\begin{figure}[b]
\centering
\subfloat[\label{fig:distribution:EUR}\EUR network]{
\includegraphics[width=\plotwidth]{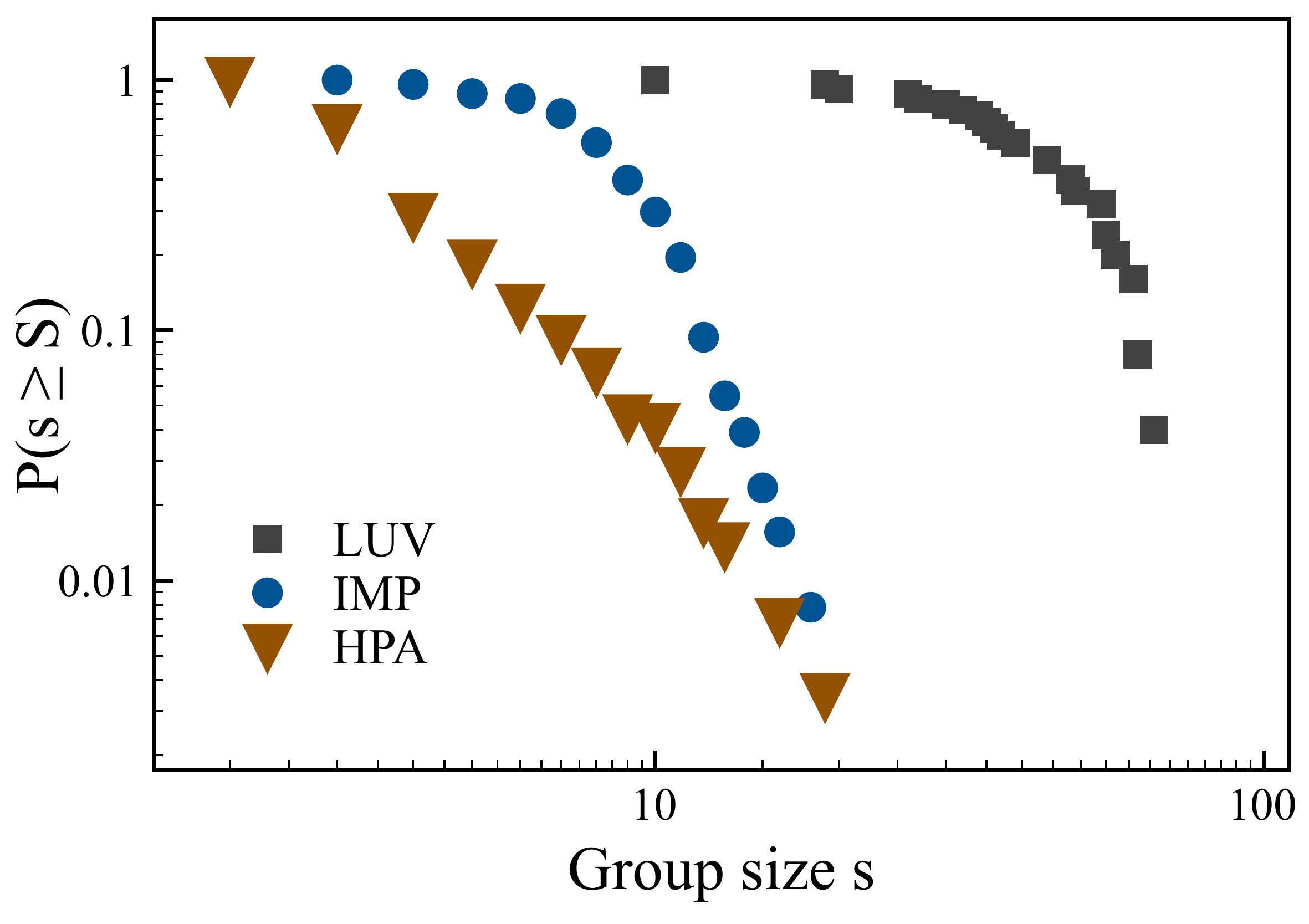}}
\subfloat[\label{fig:distribution:JVX}\JVX network]{
\includegraphics[width=\plotwidth]{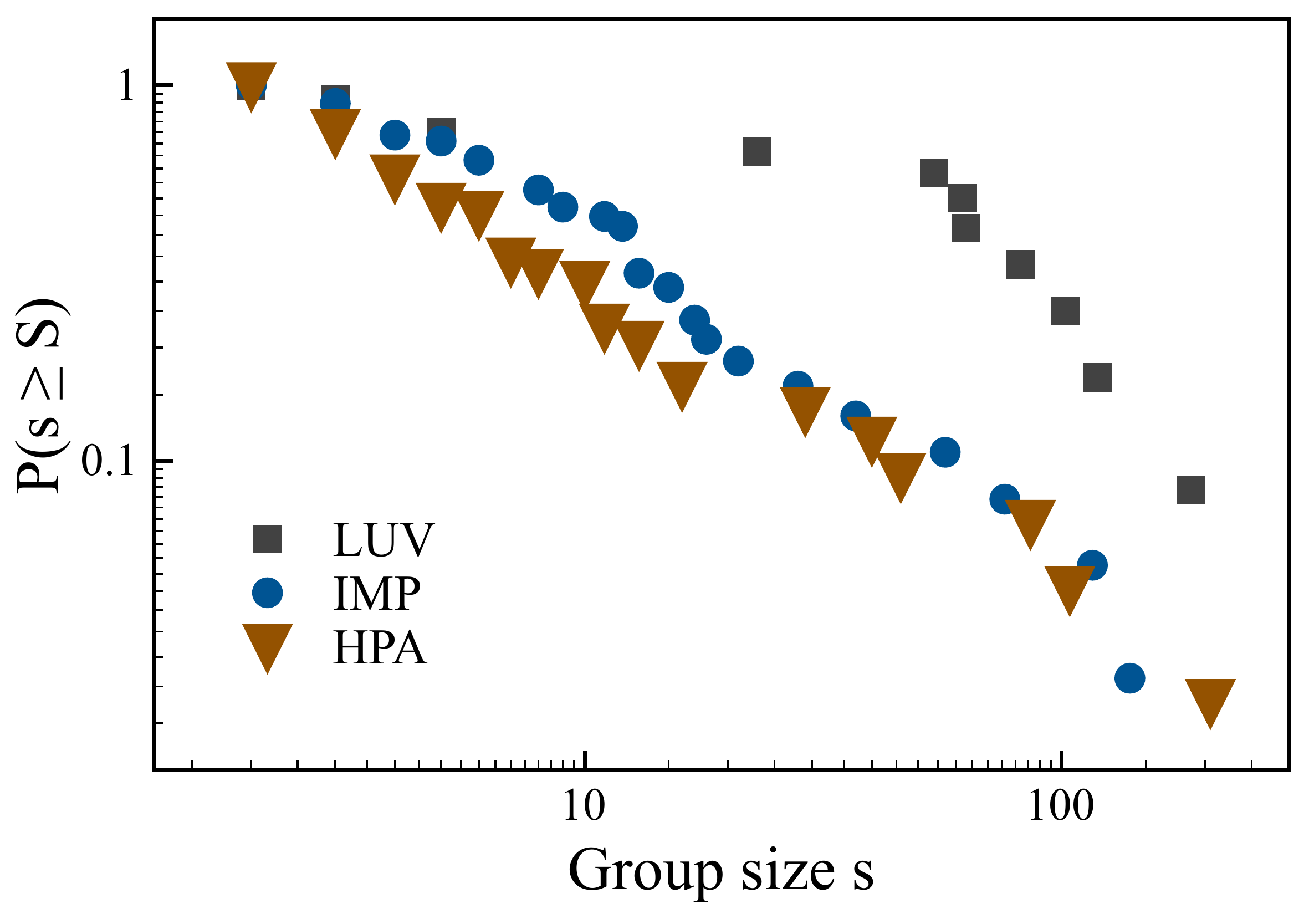}}
\caption{\label{fig:distribution}Cumulative size distributions of the group hierarchies with peak \LOGL revealed over $100$ runs.}
\end{figure}

Note that the actual performance of approaches in practical applications cannot be reliably estimated from~\tblref{realworld:likelihood}, Nevertheless, we stress that the revealed group structures differ considerably between the approaches. \figref{distribution} shows the cumulative size distributions of the groups revealed in two real-world networks (see~\tblref{realworld}). These are power-law in the case of \HPA algorithm, which is consistent with other analyses in the literature~\cite{CNM04,PDFV05}, however, the distributions in~\figref{distribution:EUR} do not appear to be heavy-tailed in the case of a community detection approach (i.e., \IMP algorithm). On the other hand, due to a resolution limit problem~\cite{FB07}, optimization of modularity~\cite{NG04} clearly favors larger groups (i.e., \LUV algorithm).

\begin{figure}[tb]
\centering
\subfloat[\label{fig:JNG:network}\JNG network]{
\includegraphics[width=\smallwidth]{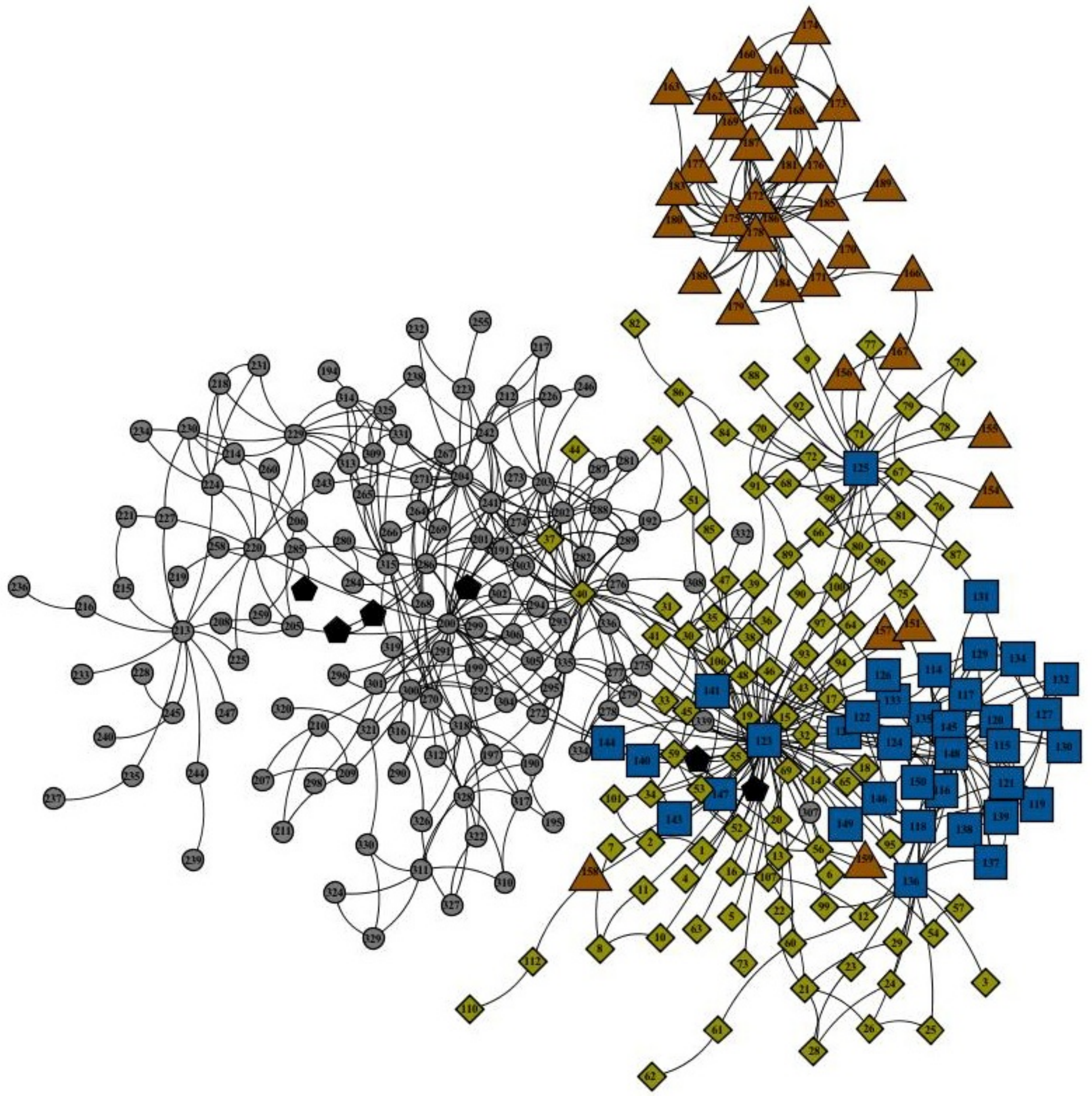}}
\subfloat[\label{fig:JNG:hierarchy:imp}Revealed with \IMP]{
\includegraphics[width=\smallwidth]{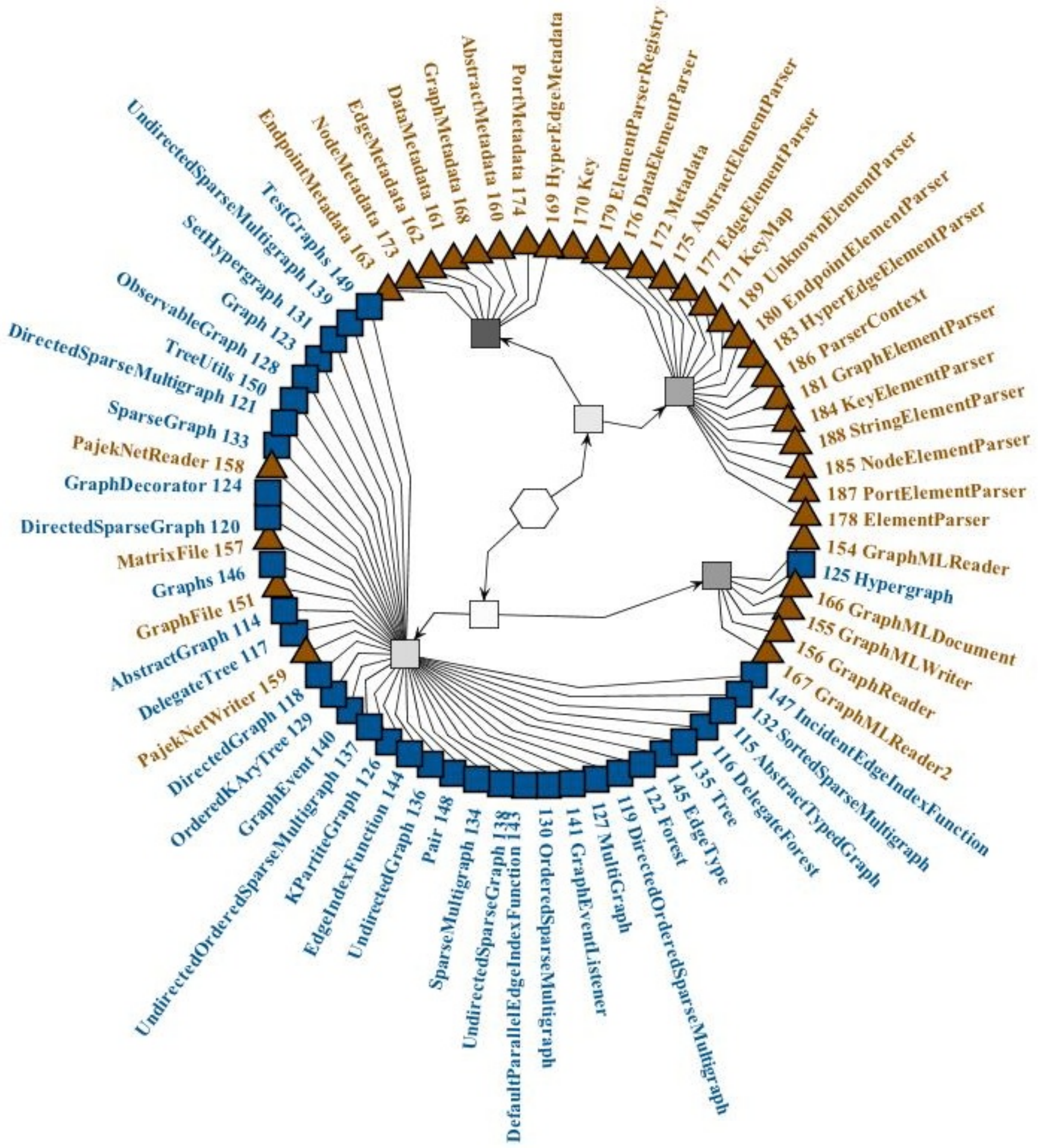}}
\subfloat[\label{fig:JNG:hierarchy:hpa}Revealed with \HPA]{
\includegraphics[width=\figurewidth]{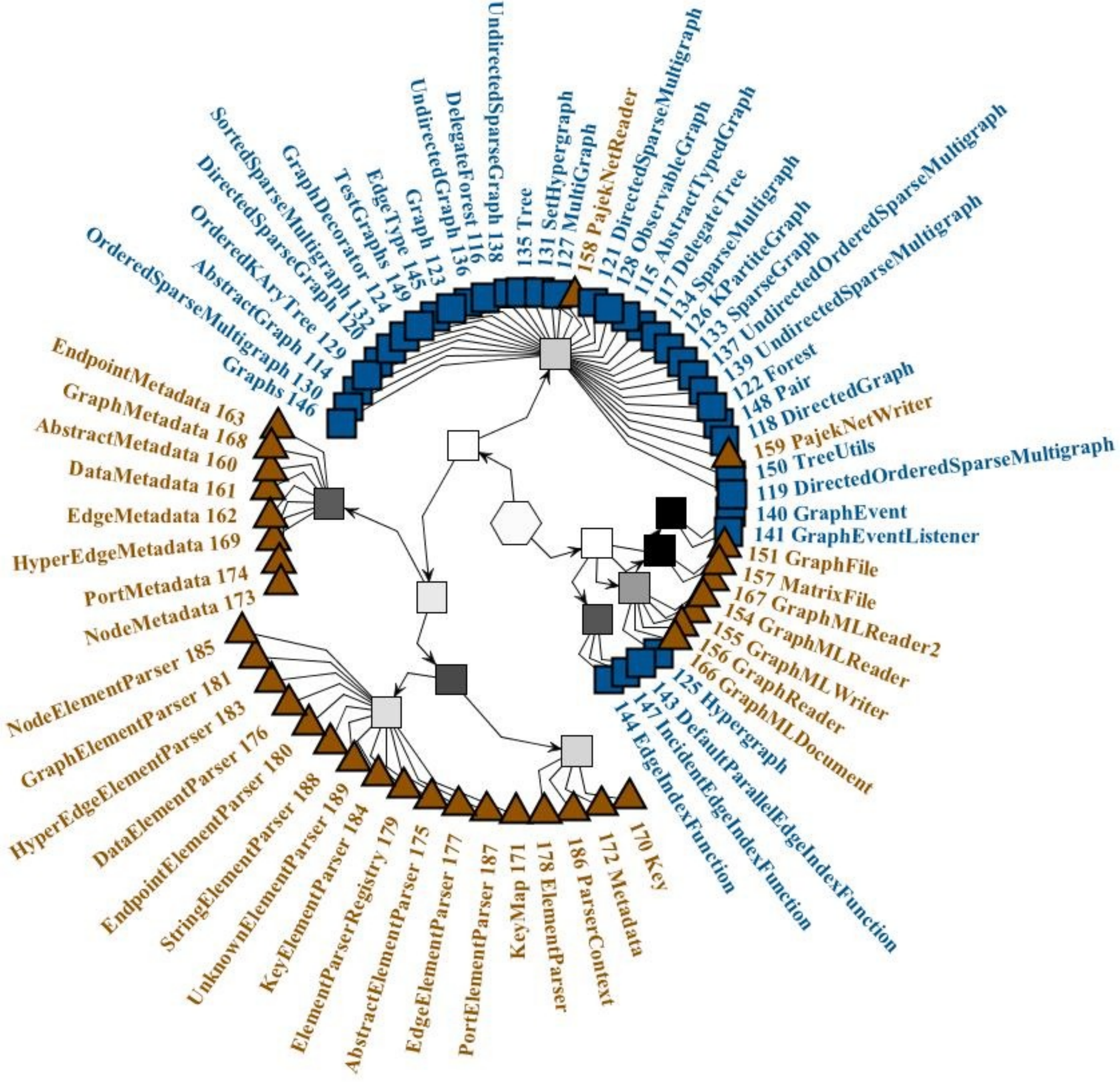}}
\caption{\label{fig:JNG}Group hierarchies of \JNG network with (b)~\logl{-500.9} and (c)~\logl{-445.3}, where node shapes correspond to high-level packages of JUNG network library, i.e., \texttt{graph} ({\color{blue} squares}), \texttt{io} ({\color{mocha} triangles}), \texttt{algorithms} ({\color{asparagus} diamonds}), \texttt{visualization}~({\color{gray} circles}) and other~({\color{black} pentagons}).  (Shades of the inner nodes of the hierarchy are proportional to probabilities $\probability$.)}
\end{figure}

We also analyze the difference between the proposed \HPA algorithm and a state-of-the-art community detection approach (i.e., \IMP algorithm) on a \JNG software network~\cite{SB11s} (see~\tblref{realworld}). Here nodes correspond to software classes, while the characteristic groups of nodes can be related to packages of the corresponding software library~\cite{SB11s,SB12s}. \figref{JNG} shows the revealed group hierarchies for only a part of \JNG network (to ease the comprehension). Both approaches reveal dense groups of nodes that correspond to graph implementations and parsers, I/O functionality and other (e.g., \figref{JNG:hierarchy:imp}). However, \HPA algorithm also further partitions, e.g., graph parsers into implementations and meta classes (i.e., \figref{JNG:hierarchy:hpa}, bottom).

\subsubsection{Network link prediction}
Last, we compare \HPA algorithm with the best community detection approach above (i.e., \IMP algorithm) on link prediction task. We consider eight real-world networks from \tblref{realworld}, where $5\%$ of the links have been removed to represent positive examples (independently, for each network realization). Approximately the same number of disconnected pairs of nodes were selected for negative examples. Next, the algorithms were applied to each realization of the reduced network, revealing a group hierarchy $\hierarchy$. The probability of a link between a pair of nodes was predicted as $\probability_i$, where $\inner_i\in\hierarchy$ is the root of the smallest sub-hierarchy including both of the nodes (see~\secref{algorithm:hpa}). Results are shown in the form of Area Under the ROC Curve~\cite{HM82} (\AUC), where higher values are better, \AUC$\in[0,1]$.

\begin{table}[htb]
\centering
\caption{\label{tbl:realworld:prediction}Mean \AUC for link prediction task over $100$ network realizations (see text for details).}
\begin{tabular}{ccccc} \HLINE
Network & $|\neighbors_i|\cdot|\neighbors_j|$ & $|\neighbors_i\cap\neighbors_j|$ & \IMP & \HPA \\\TLINE
\AFL & {$0.222$} & {$\mathit{0.817}$} & {$\mathbf{0.804}$} & {$\mathbf{0.799}$} \\
\NSC & {$0.640$} & {$\mathit{0.956}$} & {$\mathbf{0.934}$} & {$0.880$} \\
\PLB & {$0.646$} & {$\mathit{0.862}$} & {$\mathbf{0.763}$} & {$\mathbf{0.762}$} \\
\EUR & {$0.356$} & {$0.530$} & {$\mathbf{0.745}$} & {$0.700$} \\
\CLT & {$0.779$} & {$\mathit{0.826}$} & {$0.724$} & {$\mathbf{0.766}$} \\
\MNE & {$0.812$} & {$\mathit{0.920}$} & {$0.631$} & {$\mathbf{0.641}$} \\
\SWC & {$0.564$} & {$0.290$} & {$0.578$} & {$\mathbf{0.699}$} \\
\SCI & {$0.456$} & {$0.481$} & {$0.649$} & {$\mathbf{0.748}$} \\\HLINE
\end{tabular}
\end{table}

General group detection does not appear to be beneficiary for link prediction in the case of dense social networks like \NSC author collaboration network and, surprisingly, very sparse \EUR road network (\tblref{realworld:prediction}). On the other hand, \HPA algorithm most accurately predicts the underlying network structure in biological, software and two-mode social networks. Thus, the proposed group hierarchy discovery could indeed be superior to the current state-of-the-art in practice. Merely for reference, \tblref{realworld:prediction} also shows two standard link prediction techniques based on preferential attachment graph model~\cite{BA99} (i.e., $|\neighbors_i|\cdot|\neighbors_j|$) and node equivalence~\cite{LW71} (i.e., $|\neighbors_i\cap\neighbors_j|$).


\section{Conclusions and future work} \label{sec:conclusions}
The paper proposes a propagation-based general group detection algorithm for large real-world networks. The algorithm, in contrast to many other approaches, requires no apriori knowledge about the structure of the network (e.g., number of groups), while its computational complexity is near ideal. Moreover, rigorous analysis on group detection, hierarchy discovery and link prediction tasks reveals that the proposed approach is at least comparable to the current state-of-the-art. The main novelty of the paper is else a simple hierarchical refinement procedure that enables straightforward discovery of different types of groups (i.e., communities and modules), while the adopted methodology could be easily transfered to other approaches.

The paper also includes a rather extensive comparison of a larger number of group detection approaches on different synthetic benchmark graphs and real-world networks. Results give some interesting observations for future work. Most notably, while different approaches can accurately solve the community detection problem~\cite{BGLL08,RB08}, there is an absence of reliable algorithms for other groups of nodes. However, the approach based on data clustering~\cite{LKC10} performs surprisingly well on module detection task. Thus, the approach could be combined with some community detection algorithm, and the group refinement procedure proposed in this paper, into a state-of-the-art hybrid approach. Next, mixture models~\cite{NL07,KN11a} that have been very popular in recent literature are a reliable approach for group detection only in networks of moderate size, mainly due to a simple optimization procedure. Also, these require the number of groups to be known apriori, which remains to be an open problem in network science~\cite{KN11a}. On the other hand, methods based on dynamical processes like~\cite{RB08,SB11d}, and the algorithm proposed in the paper, appear to be the most reliable approaches for group detection without the above weaknesses.

The comparison in the paper does not include some otherwise very prominent group detection techniques~\cite{HEPF10,NM11a} that will be included in future work. Also, the proposed group refinement procedure may fail to recognize some less clear group configurations, which remains to be thoroughly investigated.

\appendix


\section*{Acknowledgments}
The work has been supported by the Slovene Research Agency \textit{ARRS} within the research program no.\ P2-0359.


\bibliographystyle{elsarticle-num}


\end{document}